\documentclass[12pt]{article}

\usepackage{amsmath,amssymb}
\usepackage{graphicx}

\makeatletter

\renewcommand\section{\@startsection{section}{1}{\z@}
                                   {-3.5ex \@plus -1ex \@minus -.2ex}
                                   {2.3ex \@plus .2ex}
                                   {\normalfont\large\bfseries}}
\renewcommand\subsection{\@startsection{subsection}{2}{\z@}
                                   {-3.25ex\@plus -1ex \@minus -.2ex}
                                   {1.5ex \@plus .2ex}
                                   {\normalfont\normalsize\bfseries}}
\renewcommand\subsubsection{\@startsection{subsubsection}{3}{\z@}
                                   {-3.25ex\@plus -1ex \@minus -.2ex}
                                   {1.5ex \@plus .2ex}
                                   {\normalfont\normalsize\bfseries}}
\renewcommand\paragraph{\@startsection{paragraph}{4}{\z@}
                                   {3.25ex \@plus1ex \@minus.2ex}
                                   {-1em}
                                   {\normalfont\normalsize\bfseries}}

\makeatother

\newcommand{\be}{\begin{equation}}
\newcommand{\ee}{\end{equation}}
\newcommand{\bea}{\begin{eqnarray}}
\newcommand{\eea}{\end{eqnarray}}
\newcommand{\ba}{\begin{array}}
\newcommand{\ea}{\end{array}}

\newcommand{\id}{\hbox{1\kern-.27em l}}
\newcommand{\lb}{\langle}
\newcommand{\rb}{\rangle}
\newcommand{\half}{ {\textstyle \frac{1}{2}  } }

\newcommand{\al}{\alpha}
\newcommand{\ga}{\gamma}
\newcommand{\Ga}{\Gamma}
\newcommand{\bet}{\beta}

\newcommand{\ka}{\kappa}
\newcommand{\de}{\delta}

\newcommand{\ep}{\epsilon}
\newcommand{\si}{\sigma}
\newcommand{\la}{\lambda}

\newcommand{\Om}{\Omega}
\newcommand{\ze}{\zeta}

\newcommand{\La}{\Lambda}

\newcommand{\tha}{\theta}

\newcommand{\Ups}{\Upsilon}

\newcommand{\cU}{\mathcal{U}}

\newcommand{\tr}{{\rm tr}}
\newcommand{\D}{{\rm d}}
\newcommand{\pa}{\partial}
\newcommand{\rar}{\rightarrow}
\newcommand{\non}{\nonumber}

\newcommand{\Bk}{|B\rangle}
\newcommand{\Bb}{\langle B |}

\newcommand{\vack}{|0 \rangle}
\newcommand{\vacb}{\langle 0 |}

\newcommand{\SU}{\mathrm{SU}}
\newcommand{\SO}{\mathrm{SO}}
\newcommand{\U}{\mathrm{U}}

\newcommand{\ts}{\textstyle}

\newcommand{\matrto}[4]{\left( \begin{array}{cc} #1 & #2 \\ #3 & #4 
\end{array} \right) }

\newcommand{\bpart}{\bar{\partial}}

\newcommand{\ttha}{\widetilde{\theta}}
\newcommand{\tla}{\widetilde{\lambda}}
\newcommand{\tp}{\widetilde{p}}

\newcommand{\behat}{\beta'}
\newcommand{\bethat}{\widetilde{\beta'}}
\newcommand{\gahat}{\gamma'}
\newcommand{\xihat}{\xi'}
\newcommand{\gathat}{\widetilde{\gamma'}}
\newcommand{\chihat}{\chi'}
\newcommand{\phihat}{\phi'}

\begin{document}

\setcounter{footnote}{0}
\stepcounter{table}

{\small
\begin{flushright}
CERN-PH-TH/2005-039\\
{\tt hep-th/0503123}\\
\end{flushright}
}

\vspace{5mm}

\begin{center}
{\Large\sf D--Brane Boundary States in the Pure Spinor Superstring}

\vskip 5mm

Ricardo Schiappa$^{\dag,\ddag}$ and Niclas Wyllard$^{\S}$ 

\vskip 3mm

$^\dag$CAMGSD, Departamento de Matem\'atica, Instituto Superior T\'ecnico, \\
Av. Rovisco Pais 1, 1049--001 Lisboa, Portugal \\[1mm]

$^\ddag$Faculdade de Engenharia, Universidade Cat\'olica Portuguesa, \\ 
Estrada de Tala\'ide, 2635--631 Rio de Mouro, Lisboa, Portugal \\[1mm]

$^\S$Department of Physics, Theory Division, CERN, \\
1211 Geneva 23, Switzerland

\vskip 3mm 

{\tt schiappa@math.ist.utl.pt, wyllard@cern.ch}
\end{center}

\vskip 5mm 

\begin{abstract}
\noindent
We study the construction of D--brane boundary states in the pure spinor formalism for the quantisation of the superstring. This is achieved both via a direct analysis of the definition of D--brane boundary states in the pure spinor conformal field theory, as well as via comparison between standard RNS and pure spinor descriptions of the superstring. Regarding the map between RNS and pure spinor formulations of the superstring, we shed new light on the tree level zero mode saturation rule. Within the pure spinor formalism we propose an explicit expression for the D--brane boundary state in a flat spacetime background. While the non--zero mode sector mostly follows from a simple understanding of the pure spinor conformal field theory, the zero mode sector requires a deeper analysis which is one of the main points in this work. With the construction of the boundary states at hand, we give a prescription for calculating scattering amplitudes in the presence of a D--brane. Finally, we also briefly discuss the coupling to the world--volume gauge field and show that the D--brane low--energy effective action is correctly reproduced.
\end{abstract}

\eject

\tableofcontents

\bigskip

\bigskip

\setcounter{equation}{0}
\section{Introduction and Summary}

Berkovits~\cite{Berkovits:2000a} recently proposed a new approach to covariantly quantise the superstring. This formalism has some clear advantageous features as compared to other, more traditional, approaches. For instance, one virtue is that it maintains the super--Poincar\'e symmetry of the superstring manifest, while avoiding problems associated with the quantisation of the Green--Schwarz (GS) superstring (essentially, the new formalism constructs the correct ghost sector for the classically covariant GS superstring). It also circumvents some of the more problematic aspects of the Ramond--Neveu--Schwarz (RNS) formalism, in particular the complications arising from the presence of spin fields in the vertex operators for the Ramond--Ramond (RR) fields and the spacetime fermions (as all important world--sheet fields in the new formalism have integer conformal weight, never producing branch cuts on the plane). The ghost degrees of freedom in Berkovits' approach involve certain constrained spinors, the so--called pure spinors. For this reason, the new formalism is often referred to as the ``pure spinor superstring''. Although a fair amount of research has been carried out since the appearance of~\cite{Berkovits:2000a}, there are still many other aspects of the theory which one would like to understand better, such as its origin from sigma model gauge--fixing (see, \textit{e.g.}, \cite{Matone:2002, Berkovits:2005, Aisaka:2005}). Given the promise of the pure spinor formalism to by--pass the main difficulties in both GS and RNS formalisms, thus opening the way to previously unexplored aspects of superstring theory, it becomes very important to fully develop the pure spinor superstring in all its aspects. In this paper we shall take the first steps towards developing the boundary state operator formalism~\cite{Callan:1987, Callan:1988, Green:1996} (see, \textit{e.g.},~\cite{DiVecchia:1999b, DiVecchia:1999c} for reviews) in the pure spinor superstring (see~\cite{Anguelova:2003} for earlier work in this direction). 

The contents of this paper are as follows. In the next section we begin with a review of the pure spinor formalism and fix both our conventions and notation. We also show how the refined tree level zero mode saturation rule proposed recently in \cite{Berkovits:2004a} can be obtained, starting from the well--known RNS expression. Then, in section \ref{secbs}, we present our proposal for the pure spinor boundary state in a flat spacetime background, discussing at length the most complicated part of this boundary state---its zero mode sector. In section \ref{secapp} we propose a rule for calculating (tree level) scattering amplitudes in the presence of a D--brane and check that some results, which had been previously obtained using RNS methods, are correctly reproduced in the present framework. This serves as a consistency check on our proposal for the D--brane boundary state. We also discuss the coupling of the boundary state to the world--volume gauge field living on the D--brane. In particular, we check via scattering amplitude calculations that the Dirac--Born--Infeld (DBI) and Wess--Zumino (WZ) parts of the D--brane low--energy effective action are correctly reproduced. Again, this serves as a positive check on our proposal for the D--brane boundary state. Finally, in section \ref{sdisc} we summarise our results and discuss some possible directions for future research. In the appendices some more technical aspects are collected.

\setcounter{equation}{0} 
\section{Pure Spinors and the Covariant Superstring}

The covariant quantisation of the superstring using pure spinor ghosts was initiated by Berkovits in~\cite{Berkovits:2000a}. For an introductory review we refer the reader to \cite{Berkovits:2002d}. Throughout this work we shall concentrate on the pure spinor version of the superstring in a flat spacetime background. Let us note that in this paper we will sometimes find it useful to compare expressions obtained for the pure spinor superstring to analogous ones in the RNS superstring, partly because this provides additional motivation for some of our expressions and partly because most readers are more familiar with the standard RNS formalism. Even though our comparisons with RNS have various levels of detail, they are mostly at the heuristic level (possibly with the exception of the discussion in section \ref{sMes}) and should thus not be thought of as rigourous derivations. It is probably possible to make some of our statements precise, by carefully implementing the map between RNS and pure spinor descriptions of the superstring given in~\cite{Berkovits:2001a}.  However, we believe that it is more important to develop and understand the pure spinor superstring formalism at the covariant level, and that our ultimate goal is to obtain a complete understanding of this theory, independently of its RNS counterpart. Thus, we shall leave our RNS derivations to be regarded mostly as motivations for the results we obtain. Another point we would like to stress is that in this paper we use bosonisation at certain intermediate stages of our considerations. However, in the final pure spinor expressions there is no need to bosonise the variables, a fact which has been previously argued to be one of the advantageous features of the pure spinor superstring formalism.

\subsection{Type II Covariant Superstrings in a Flat Background}
\label{flat-back}

We begin by reviewing the pure spinor version of the superstring in a flat background, initiated in \cite{Berkovits:2000a}, and later developed in a long series of papers, \textit{e.g.}, \cite{Berkovits:2000b, Berkovits:2000c, Berkovits:2001a, Berkovits:2001c, Berkovits:2001d, Berkovits:2002b}. We have attempted to make the presentation reasonably self--contained; more details can be found in the references.

In the pure spinor version of the type II superstring, the world--sheet fields are $(x^{m}, \theta^{\alpha}, \widetilde{\theta}^{\widetilde{\alpha}})$---the world--sheet analogues of the ${\mathcal{N}}=2$, $d=10$ superspace variables--- together with~\cite{Siegel:1985} $( p_{\alpha}, \widetilde{p}_{\widetilde{\alpha}} )$ where $p_{\alpha}$ is the conjugate momentum to $\theta^{\alpha}$ and $\widetilde{p}_{\widetilde{\alpha}}$ is the conjugate momentum to $\widetilde{\theta}^{\widetilde{\alpha}}$. Above $m = 0, \dots, 9$ and $\alpha, \widetilde{\alpha} = 1, \dots, 16$. In the type IIA superstring, where the target space has $(1,1)$ supersymmetry, $\alpha$ and $\widetilde{\alpha}$ denote $\SO(1,9)$ Majorana--Weyl spinors of opposite chirality, while in the type IIB superstring, where the target space has $(2,0)$ supersymmetry, $\alpha$ and $\widetilde{\alpha}$ denote $\SO(1,9)$ Majorana--Weyl spinors of the same chirality (in the case of the type IIB theory we shall later drop the $\widetilde{}$ on the $\widetilde{\al}$ indices to simplify the notation). In the pure spinor formalism, the world--sheet ghost fields are $\lambda^{\alpha}$ and $\widetilde{\lambda}^{\widetilde{\alpha}}$, which are complex bosonic Weyl spinors constrained to satisfy the pure spinor conditions \cite{Berkovits:2000a}

\begin{equation}
\lambda \gamma^{m} \lambda = 0\,, 
\qquad \widetilde{\lambda} \gamma^{m} \widetilde{\lambda} = 0 \,.
\label{pure-spinor}
\end{equation}

\noindent
The pure spinor conditions (\ref{pure-spinor}) reduce the number of independent complex components of both $\la^\al$ and 
$\widetilde{\lambda}^{\widetilde{\al}}$ from sixteen to eleven. 
It is important to notice that even though $\la^\al$ and $\widetilde{\lambda}^{\widetilde{\al}}$ 
are complex, they enter holomorphically in the theory (\textit{i.e.}, their complex 
conjugates, $\bar{\la}^\al$ and $\bar{\widetilde{\lambda}}^{\widetilde{\al}}$, 
never appear in the world--sheet action). This 
means that, \textit{e.g.}, in the determination of the central charge (see below) 
the counting of degrees of freedom is the same as if 
$\la^\al$ and $\widetilde{\lambda}^{\widetilde{\al}}$ were real.
In (\ref{pure-spinor}), the gamma matrices
$\gamma^{m}$ are the $16 {\times} 16$ off--diagonal blocks 
(``Pauli matrices'') in the Weyl representation of the 
$32 {\times} 32$ ten--dimensional gamma matrices $\Gamma^{m}$. 
These two matrices are symmetric $\gamma^{m}_{\alpha\beta} =
\gamma^{m}_{\beta\alpha}$, ${\gamma^{m}}^{\alpha\beta} =
{\gamma^{m}}^{\beta\alpha}$, and satisfy

\begin{equation}
\gamma^{m}_{\alpha\beta} {\gamma^{n}}^{\beta\sigma} +
\gamma^{n}_{\alpha\beta} {\gamma^{m}}^{\beta\sigma} = 2 \eta^{mn}
\delta_{\alpha}^{\sigma}\,,
\end{equation}

\noindent
so that $\left\{ \Gamma^{m}, \Gamma^{n} \right\} = 2 \eta^{mn}
\id_{32}$. 

In a flat spacetime background, $\theta^{\alpha}$,
$p_{\alpha}$ and $\lambda^{\alpha}$ are left--moving (holomorphic), while
$\widetilde{\theta}^{\widetilde{\alpha}}$, 
$\widetilde{p}_{\widetilde{\alpha}}$ and 
$\widetilde{\lambda}^{\widetilde{\alpha}}$  
are right--moving\footnote{In many papers 
$\,\widehat{}$ (hat) is used instead of $\,\widetilde{}$ (tilde) to denote 
the right--moving variables.} 
(anti--holomorphic). In the following we mostly only 
write the expressions involving the left--movers. 
The formul\ae{} for the right--moving fields will look essentially
the same. 
The world--sheet action in a flat background is (in units where 
$\alpha'=2$) \cite{Berkovits:2000a}

\begin{equation}
S = \frac{1}{2\pi} \int d^{2} z\ \left( \half \partial x^{m}
\bar{\partial} x_{m} + p_{\alpha} \bar{\partial} \theta^{\alpha} \right)
+S_{\lambda} \,,
\label{flat-action}
\end{equation}

\noindent
where $S_{\lambda}$ is the action for
the $\lambda^{\alpha}$ ghosts. 
The free fields $x^m$, $\tha^\al$ and $p_\al$ have the standard OPE's,

\begin{equation}
x^{m} (y, \bar{y}) x^{n} (z, \bar{z}) \sim -  \eta^{mn} \log
\left| y - z \right|^{2}, \quad p_{\alpha} (y) \theta^{\beta} (z) \sim
\frac{\delta_{\alpha}^{\beta}}{y-z}\,. 
\end{equation}

\noindent
Because of the pure spinor constraint (\ref{pure-spinor}) the conjugate 
momenta to  $\lambda^{\alpha}$, $w_{\alpha}$, is only defined modulo 
the transformation $w_\al \rar w_\al + (\ga^m \la)_\al \La_m $. 
Although one could write an action for the ghost fields involving 
$w_\al \bar{\pa} \la^\al$ this form would be slightly deceiving 
since the pure spinor constraint implies, for example, that the OPE between 
$w_{\al}$ and $\la^{\beta}$ is not the canonical one, \textit{cf.} (\ref{wlOPE}).  
Nevertheless, it is still possible to work with this form of the ghost action 
provided one carefully takes into account the invariance under the 
transformation of $w_\al$ mentioned above 
(see, \textit{e.g.}, \cite{Chesterman:2002} for a discussion).
Another possible way around the problem of the non--covariance of $w_\al$ 
is to relax the pure spinor constraint. This approach has been pursued 
in a number of works 
(see, \textit{e.g.}, \cite{Grassi:2003b, Aisaka:2003, Grassi:2004} and references 
therein) but will not be further discussed in this paper. 
Yet another way to write down an explicit form for the ghost action, 
$S_{\lambda}$, in terms of free fields is to Wick rotate and to temporarily 
break the manifest $\SO(10)$ Lorentz invariance to 
$\U(5) \simeq \SU(5) {\times} \U(1)$ \cite{Berkovits:2000a}. 
As explained in \cite{Berkovits:2001a},  $\U(5)$ is 
the maximal subgroup of $\SO(10)$ which leaves the pure spinor constraint 
invariant.

As we have reviewed in appendix \ref{u5}, an explicit parameterisation of $\la^\al$ 
satisfying the constraint (\ref{pure-spinor}) is \cite{Berkovits:2001a} 

\begin{equation} \label{lapar}
\lambda^{+} = e^{s}, \quad \lambda_{ab} = u_{ab}, \quad \lambda^{a} = 
{\ts \frac{1}{8}} e^{-s} \epsilon^{abcde} u_{bc} u_{de}\,,
\end{equation}

\noindent
where $a = 1, \ldots, 5$, and $u_{ab} = - u_{ba}$ are ten complex variables
(which together with their conjugate momenta, $v^{ab}$, parameterise 
the $\SO(10)/\U(5)$ coset). The parameterisation (\ref{lapar}) is 
well defined as long as $\lambda^{+} \not = 0$. 

Using these ``$\U(5)$ variables'', the ghost action can be explicitly 
written as\footnote{Note that the integrand equals  
$-w_{\al}\bar{\pa}\la^{\al}$ as can be seen using (\ref{w}) with 
a suitable choice of the parameter $a$, together with the definitions 
of $\behat$ and $\gahat$ given below.} 

\begin{equation} \label{u5action}
S_{\lambda} = \frac{1}{2\pi}\int d^{2} z\ \left( v^{ab} \bar{\partial}
u_{ab} + \behat \bar{\partial} \gahat \right),
\end{equation}

\noindent
where $\behat$ has conformal weight 1 (let us stress that this $\bet\ga$--system is not the standard one appearing in the RNS formulation). It turns out that it is convenient to bosonise $(\behat,\gahat)$ according to 
$\behat \cong e^{-\phihat +\chihat}\pa\chihat$ and 
$\gahat \cong e^{\phihat - \chihat}$ and define 
$s = \half(\chihat - \phihat)$ and $t=\chihat+\phihat$. 
It then follows that $\gahat = e^{-2s}$. 
As we shall later see in greater detail, this $s$ is the same 
as the one appearing in (\ref{lapar}). 
The free fields $s$, $u_{ab}$, and their conjugate momenta $t$, $v^{ab}$, satisfy the free field OPE's,

\begin{equation} \label{stOPE}
t(y) s(z) \sim \log \left( y - z \right), \quad v^{ab} (y) u_{cd} (z) \sim
-\frac{\delta^{ab}_{cd} }{y-z}\,,
\end{equation}

\noindent
where $\de^{ab}_{cd} = \frac{1}{2}(\de^a_c\de^b_d - \de^a_d\de^b_c)$.

The holomorphic stress tensor in the pure spinor superstring is

\begin{equation}
T = - \half \partial x^{m} \partial x_{m} - p_{\alpha} \partial
\theta^{\alpha} + T_{\lambda} \,,
\end{equation}

\noindent
where $T_{\lambda}$ is the stress tensor for the 
$\lambda^{\alpha}$ ghosts. In terms of the free $\U(5)$ variables, 
the ghost stress tensor can be written as 
(later we show that $\partial t \partial s +
\partial^{2} s = -\behat \pa \gahat$)

\begin{equation} \label{ghosT}
T_{\lambda} = -v^{ab} \partial u_{ab} + \partial t \partial s +
\partial^{2} s \,.
\end{equation}

\noindent
Using this expression it can be readily checked that the ghost CFT has $c=22$. 
If one recalls that the $x^m$ CFT has the standard
$c=10$ central charge, while the $(p,\theta)$ CFT has central charge $c=-32$,
the total central charge vanishes as required.

Even though the OPE between $w_\al$ and $\la^\al$ is not manifestly $\SO(10)$ 
covariant, one can nevertheless explicitly construct $\SO(10)$ 
Lorentz currents for the ghosts as~\cite{Berkovits:2000a}

\begin{equation}
N^{mn} = \half w \gamma^{mn} \lambda \,.
\end{equation}

\noindent
As shown in~\cite{Berkovits:2000a}, and discussed at greater length 
in appendix \ref{u5}, the OPE's involving $N_{mn}$ and $\la^\al$ 
have the manifestly $\SO(10)$ covariant form 

\begin{eqnarray} \label{OPE-N}
N^{mn} (y) \lambda^{\alpha} (z) &\sim& \frac{1}{2}\ \frac{1}{y - z}
{\left( \gamma^{mn} \right)^{\alpha}}_{\beta} \, \lambda^{\beta} (z)\,, \\
N^{p q } (y) N^{mn} (z) &\sim& \frac{ 
\eta^{p m} N^{q  n} (z) {-} \eta^{q m} N^{p n} (z)
 - (m\leftrightarrow n)
}{y - z} -
3 \frac{\eta^{p n} \eta^{q m} {-} \eta^{p m}
\eta^{q n}}{(y - z)^{2}} \,. \non
\end{eqnarray}

\noindent
From this expression we see that the ghost Lorentz currents $N^{mn}$ form 
a $\SO(10)$ current algebra with level $k=-3$. In comparison, the OPE's involving the $(p,\tha)$ Lorentz current, $M^{mn} = -\frac{1}{2} p \gamma^{mn} \theta$, take the form

\begin{eqnarray} \label{OPE-M}
M^{mn} (y) \theta^{\alpha} (z) &\sim&  \frac{1}{2}\ \frac{1}{y - z}
{\left( \gamma^{mn} \right)^{\alpha}}_{\beta} \, \theta^{\beta} (z)\,, \\
M^{p q} (y) M^{mn} (z) &\sim& 
\frac{ \eta^{p m } M^{q n} (z) {-} \eta^{ q m }
M^{p  n} (z) - (m\leftrightarrow n)
}{y - z} +
4 \frac{\eta^{p n} \eta^{q m} {-} \eta^{p m}
\eta^{q n}}{(y - z)^{2}} \,. \non
\end{eqnarray}

\noindent
Thus the $M^{mn}$'s also form a $\SO(10)$ current
algebra, this time at level $k=4$. The total Lorentz current 
$L_{mn} = -\frac{1}{2} p \gamma^{mn} \theta + N^{mn}$ satisfies the OPE

\begin{eqnarray}
\!\!\!\!\!\!\!\!\!\! 
L^{p q} (y) L^{mn} (z) \!\!\! &\sim& \!\!\! \frac{ \eta^{p m}
L^{q n}(z)  {-} \eta^{q m}
L^{p  n}(z) - (m\leftrightarrow n) }{y - z} 
{+} \frac{\eta^{p n} 
\eta^{q m} {-} \eta^{p m}
\eta^{q n}}{(y - z)^{2}}
\label{OPE-LL}
\end{eqnarray}

\noindent
and thus forms a current algebra with level $k=1$ as expected from 
comparison with the RNS formalism \cite{Berkovits:2000a}, 
where the $L^{mn} = -\Psi^m \Psi^n$ satisfy (\ref{OPE-LL}).

As should be clear from the above discussion, the OPE's of the fields 
$\lambda^{\alpha}$ and the currents
$N^{mn}$ are manifestly covariant (even though they were computed 
starting from the non--covariant free ghost action). Moreover,
although one can in principle write down pure spinor vertex operators 
in terms of the free variables, it turns out that the requirement 
of super--Poincar\'e
covariance implies that the free variables $\left\{ s, t, u_{ab}, v^{ab}
\right\}$ can only appear in the Lorentz covariant combinations of $\left\{
\lambda^{\alpha}, N^{mn}, \partial h \right\}$, where $h$ is a 
Lorentz scalar defined through

\begin{equation}
\partial h = \half w_{\alpha} \lambda^{\alpha}\,.
\end{equation}

\noindent
As was the case for $N^{mn}$, even though $\pa h$ contains 
the ``non--covariant'' 
quantity $w_\al$, the OPE's involving $h$, $N^{mn}$ and $\la^\al$ 
are manifestly Lorentz covariant. Observe that both $N^{mn}$ and $\pa h$ 
are invariant under the transformation $w_\al \rar w_\al 
+ (\ga^{m}\la)_\al \La_m$, because of the pure spinor constraint on $\la^\al$. 
The Lorentz scalar $\pa h$ has 
no singularities with the Lorentz currents $N^{mn}$ and satisfies the  OPE's

\begin{equation}\label{extra1}
h(y) h(z) \sim - \log \left( y - z \right), \quad \partial h(y)
\lambda^{\alpha} (z) \sim \frac{1}{2}\ \frac{1}{y-z} \lambda^{\alpha} (z)\,.
\end{equation}

\noindent
It is possible to show that the operator $J= 2 \oint \partial h$
is the ghost number charge, implying that  $\lambda^\al$ has ghost number one
as it should have.

Furthermore, using the covariant fields $N^{mn}$ and $\pa h$, the ghost stress tensor can be written in a manifestly Lorentz invariant way as \cite{Berkovits:2001a}

\begin{equation}\label{covst}
T_{N,\pa h} 
= - {\ts \frac{1}{20}} N_{mn} N^{mn} - \half \left
( \partial h \right)^{2} + 2 \partial^{2} h \,.
\end{equation}

\noindent
By using various normal ordering 
rearrangements, one can check that the stress tensor (\ref{covst}) 
indeed reduces to the stress tensor (\ref{ghosT}) written in terms 
of the $\U(5)$ variables, $s$, $t$, $u_{ab}$ and $v^{ab}$ 
(see appendix \ref{u5} for further details). 
Let us analyse the ghost stress tensor (\ref{covst}) 
in more detail. The first piece
involves the ghost Lorentz currents, $N^{mn}$, and is a Sugawara
construction for a $\SO(10)$ WZNW  model at level $k=-3$. 
Indeed, recalling that 
the dual Coxeter number of $\SO(10)$ is $g=8$, we find\footnote{Due to our 
normalisation of the $N N$ OPE in (\ref{OPE-N}) the prefactor in front of 
$N_{mn}N^{mn}$  in (\ref{covst}) is unconventional. To obtain the usual 
$+\frac{1}{10}$ one would have to rescale the currents $N^{mn}$.} $2(g+k) = 10$. 
The second piece refers to a Coulomb gas, with a background
charge of $Q=4$. Using standard formul\ae{} one finds that the central charge 
of the ghost Lorentz currents is

\begin{equation}
c = \frac{k \dim \SO(10)}{k+g} = \frac{(-3)(45)}{-3+8} = -27\,,
\end{equation}

\noindent
while the central charge of the Coulomb gas field is $c = 1 + 3 Q^{2} =
49$, so that $-27+49=22$ as expected. The expression for the stress tensor 
(\ref{covst}) also allows one to write explicitly covariant expressions 
for the ghost action, albeit in the complicated form as a sum of a WZNW 
and a Coulomb gas action.

To obtain a better understanding of the form of the pure spinor stress tensor 
we note that it is also possible to rewrite the $(p, \tha)$ stress tensor 
in the form

\begin{equation}\label{covst2}
T_{M,\pa g} = - {\ts \frac{1}{48}} M_{mn} M^{mn} + \half \left
( \partial g \right)^{2} - 2 \partial^{2} g \,,
\end{equation}

\noindent
where, as above, $M^{mn} = -\frac{1}{2}p\ga^{mn}\tha$ and satisfies the 
OPE's (\ref{OPE-M}), while the Lorentz scalar 
$\pa g = \frac{1}{4}p_\al \tha^\al$ has no singularities with $M^{mn}$ 
and satisfies the OPE's

\begin{equation}\label{extra2}
g(y) g(z) \sim \log \left( y - z \right), \quad \partial g(y)
\tha^{\alpha} (z) \sim -\frac{1}{4}\ \frac{1}{y-z} \tha^{\alpha} (z)\,.
\end{equation}

\noindent
In the $(p, \tha)$ stress tensor, the spacetime fermion Lorentz currents, $M^{mn}$, appear as a Sugawara construction of a $\SO(10)$ WZNW model at level $k=4$, and the Coulomb gas piece has background charge $Q=4$. Further observe that the operator $K= 4 \oint \partial g$ is the spacetime fermion number charge, implying that  
$\theta^\al$ has fermion number $-1$. 
In the form (\ref{covst2}) the central charge is calculated 
as $4\cdot 45/12 + (1-3\cdot4^2) = 15 - 47 = -32$. 
It is a straightforward but tedious exercise to 
show that after using various normal ordering rearrangements the  
stress tensor (\ref{covst2}) indeed reduces 
to $T_{p\tha} = -p_\al \pa \tha^\al$. 

We have thus learned that the full stress tensor (excluding the $\partial x$ part) 
can be written as a sum of two WZNW and two Coulomb gas pieces.  
In~\cite{Grassi:2003, Grassi:2004} it has also 
been noted that the pure spinor superstring (in the extended formulation 
where the pure spinor constraint is relaxed) can be formulated 
in terms of WZNW models.

The above discussion can be summarised in the following table, 
containing the conformal weights and ghost number assignments of the
world--sheet fields:

\vspace*{-2pt}
\begin{center}
\begin{tabular}{|c|c|c|}
\hline
\textsf{Field} & \textsf{Conformal Dimension} & \textsf{Ghost Number} \\ 
\hline
$\partial x^{m}$, $p_{\al}$ & $1$ & $0$ \\ \hline
$\theta^{\alpha}$ & $0$ & $0$ \\ \hline
$\lambda^{\alpha}$ & $0$ & $1$ \\ \hline
$M^{mn}$, $N^{mn}$  & $1$ & $0$ \\ \hline
$\pa g$, $\pa h$  & $1$* & $0$ \\ \hline
\end{tabular}
\\[8pt]
{\footnotesize {\bf\sf Table 1:} Conformal dimensions and ghost numbers of the 
various fields. The * indicates that \\ $\pa g$ and $\pa h$ 
do not have honest conformal dimensions because of their background charges.}
\\[5pt]
\end{center}

\noindent
These properties can be derived using both the free field form of the 
above stress tensors or the WNZW/Coulomb gas one, but depending on what 
one wants to calculate it is usually easier to use one of the 
two (equivalent) versions.

The physical states of the superstring in the pure spinor formalism are
obtained from vertex operators in the cohomology of the 
left-- and right--moving BRST operators \cite{Berkovits:2000a}

\begin{equation} \label{QBRST}
Q = \oint \lambda^{\alpha} d_{\alpha}\,, \qquad \widetilde{Q} = \oint
\widetilde{\lambda}^{\widetilde{\alpha}}
\widetilde{d}_{\widetilde{\alpha}}\,,
\end{equation}

\noindent
where (the definition of $\widetilde{d}_{\widetilde{\al}}$ is completely analogous) 

\begin{equation}
d_{\alpha} = p_{\alpha} - \half \left( \gamma^{m} \theta \right)_{\alpha}
\partial x_{m} - {\ts \frac{1}{8} } 
\left( \gamma^{m} \theta \right)_{\alpha}
\left( \theta \gamma_{m} \partial \theta \right) .
\end{equation}

\noindent
The world--sheet field $d_{\alpha}$ satisfies the OPE's\footnote{In order to
verify the first of these OPE's, it is crucial to use the gamma matrix
identity $\left( \gamma^{m} \theta \right)_{\alpha} \left( \gamma_{m}
\xi \right)_{\beta} = \frac{1}{2} \left[ \left( \gamma^{m} \theta
\right)_{\alpha} \left( \gamma_{m} \xi \right)_{\beta} - \left( 
\gamma^{m} \theta \right)_{\beta} \left( \gamma_{m} \xi
\right)_{\alpha} \right] - \frac{1}{2} \gamma^{m}_{\alpha\beta} 
\left( \theta \gamma_{m} \xi \right)$.}

\bea
&& d_{\alpha} (y) d_{\beta} (z) \sim - \frac{1}{y - z}
\gamma_{\alpha\beta}^{m} \Pi_{m} (z)\,, \quad
d_{\alpha} (y) \Pi^{m} (z) \sim \frac{1}{y - z}
\left( \gamma^{m} \partial \theta \right)_{\alpha} (z)\,,\non \\
&& d_{\al}(y)\pa \theta^\beta(z) \sim \frac{1}{(y-z)^2}\de^\beta_\al \,,
\label{OPE-dd}
\eea

\noindent
where

\begin{equation}
\Pi^{m} = \partial x^{m} + \half \theta \gamma^{m} \partial
\theta \,.
\end{equation}

\noindent
Using the above OPE's it is easy to check that 
the BRST operator is nilpotent 
(due to the pure spinor condition (\ref{pure-spinor})), so that $Q^2 = 0 = {\widetilde{Q}}^2$.

The spacetime supersymmetry generator is

\begin{equation} \label{SuzyQ}
q_{\alpha} = \oint \left( p_{\alpha} + \half \left( \gamma^{m} \theta
\right)_{\alpha} \partial x_{m} + {\ts \frac{1}{24} }
\left( \gamma^{m} \theta
\right)_{\alpha} \left( \theta \gamma_{m} \partial \theta \right) \right),
\end{equation}

\noindent
and  satisfies the anticommutation relation

\begin{equation}
\left\{ q_{\alpha}, q_{\beta} \right\} =  \gamma^{m}_{\alpha\beta} \oint
\partial x_{m}\,.
\end{equation}

\noindent
It is straightforward to check that the BRST operator as well as the 
world--sheet fields $d_{\al}$, $\Pi^m$ are supersymmetric, 
\textit{i.e.}, they (anti)commute with $q_{\al}$.

For an arbitrary superfield $\Phi (x,\theta)$, one has that 
$[ \oint d_{\alpha} \,, \Phi (x,\theta) \} = D_{\alpha} \Phi (x,\theta)$ 
where\footnote{We refer the reader to appendix \ref{appFlat} for our superspace 
conventions.}

\begin{equation}
D_{\alpha} = \frac{\partial}{\partial \theta^{\alpha}} + \frac{1}{2} \left( 
\gamma^{m} \theta \right)_{\alpha} \frac{\partial}{\partial x^{m}}\,.
\end{equation}

\noindent
Thus, the world--sheet field $d_{\alpha}$ corresponds to the spacetime 
supersymmetric covariant derivative. Similarly, one has  
$[q_{\alpha} \, \Phi (x,\theta)\} = 
\mathcal{Q}_{\alpha} \Phi (x,\theta)$, where 

\begin{equation}
\mathcal{Q}_{\alpha} = \frac{\partial}{\partial \theta^{\alpha}} -
\frac{1}{2} \left( \gamma^{m} \theta \right)_{\alpha}
\frac{\partial}{\partial x^{m}}\,.
\end{equation}

Vertex operators in flat space were constructed in \cite{Berkovits:2000a,
Berkovits:2000b, Berkovits:2002b, Cornalba:2002, Grassi:2004b}. 
The unintegrated vertex operator
for the massless {\em open} superstring state 
is $U = \lambda^{\alpha} A_{\alpha}
(x,\theta)$, where $A_{\al}(x,\tha)$ is a superfield. 
Since $\{Q, A_\beta (x,\theta)\} 
= \lambda^{\alpha} D_{\alpha} A_\beta (x,\theta)$, the physical requirement 
$\{Q, U\} = 0$ implies the ten--dimensional (superspace) 
Yang--Mills equations of motion 
(using that for pure spinors $\la^\al \la^\beta 
= \frac{1}{1920} \ga_{m_1 \cdots m_5}^{\al\beta} 
 \la \ga^{m_1\cdots m_5}\la$)

\begin{equation}
\gamma^{\alpha\beta}_{m_{1} \cdots m_{5}} D_{\alpha} A_{\beta} = 0\,. 
\end{equation}

\noindent
Hence $A_{\al}$ is the spinor
superfield potential for super Yang--Mills theory (see 
appendix \ref{appFlat} for more details). Moreover, the BRST  
invariance $\delta U = [Q, \Omega]$ implies the Yang--Mills gauge 
transformations $\delta A_{\alpha} = D_{\alpha} \Omega$. 
Choosing an appropriate gauge and restricting to constant field strengths, 
one can write (see also appendix \ref{appFlat})

\begin{equation} \label{openU}
U =  \half \left( \lambda \gamma^{m} \theta \right) a_{m} (x) 
+ {\ts \frac{1}{3} } \left
( \lambda \gamma^{m} \theta \right) \left( \theta \gamma_{m} \xi
\right)
-
{\ts \frac{1}{32} } \left( \lambda \gamma_{\sigma} \theta \right) \left( \theta
\gamma^{\sigma m n} \theta \right) f_{mn}  ,
\end{equation}

\noindent
where $a_{m} (x)$ is the gluon with constant field strength $f_{mn}$,
and $\xi^{\alpha}$ is the constant gluino. For non--constant fields there 
will be additional terms in (\ref{openU}).

Integrated vertex operators, $\oint V$, are defined through $\left[ Q, V
\right] = \partial U$, and are thus manifestly BRST invariant. For the
massless states of the open superstring,

\begin{equation}
V = \Pi^{m} A_{m} (x, \theta) + \partial \theta^{\alpha} A_{\alpha} (x,
\theta) + d_{\alpha} W^{\alpha} (x, \theta) + \half N^{mn}
F_{mn} (x, \theta) \,.
\end{equation}

\noindent
The definitions of the superfields $A_{m}$, 
$W^{\alpha}$ and $F_{mn}$ can be found in appendix \ref{appFlat}. To lowest
order in the component fields, and in the same gauge as before,

\begin{equation} \label{fint}
V = a_{m} (x) \partial x^{m} + \half f_{mn} \left( 
-\half p \gamma^{mn} \theta + N^{mn} \right) + \xi^{\alpha}
q_{\alpha}.
\end{equation}

\noindent
For the closed superstring the story is similar (see \cite{Grassi:2004b} 
for an extensive recent discussion). The
massless unintegrated vertex operator is $U = \lambda^{\alpha}
\widetilde{\lambda}^{\widetilde{\beta}} A_{\alpha\widetilde{\beta}}
(x,\theta,\widetilde{\theta})$, where $A_{\alpha\widetilde{\beta}}$ is a
$\mathcal{N}=2$, $d=10$ bispinor superfield.
The physical state conditions are\footnote{In this paper we shall
ignore the subtleties associated with zero momentum states.} 
$[Q, U] = 0 = [\widetilde{Q}, U ]$
implying the (linearised) supergravity equations of motion, 
and the gauge invariance
is $\delta U = \{Q, \widetilde{\Omega}\} + \{\widetilde{Q}, \Omega\}$ with
$\{\widetilde{Q}, \widetilde{\Omega}\} = 0 = \{Q, \Omega\}$. 
In order to construct the component forms of the massless 
closed string vertex operators, the simplest way to proceed is to use the
fact that these operators can be understood as the left--right product of
the open string vertex operators that we described above. The vertex
operator,

\begin{equation} \label{UNS}
\lambda A \widetilde{\lambda} = \left( \lambda \gamma^{m} \theta \right)
\zeta_{mn} (x) ( \widetilde{\lambda} \gamma^{n} \widetilde{\theta})\,, 
\end{equation}

\noindent
with the gauge choice $k^m \zeta_{mn}=0$, $k^2 \, \zeta_{mn}=0$, 
describes a graviton when 
$\zeta_{mn} = h_{mn} = h_{nm}$ with $\eta^{mn} h_{mn}=0$.
It describes a $B$--field when $\zeta_{mn} = B_{mn} = -B_{nm}$ 
and it describes the dilaton when $\zeta_{mn}(x) = \zeta(x) \, \epsilon_{mn}$ 
and

\begin{equation}
\epsilon_{mn} = \frac{1}{\sqrt{d-2}} \left( \eta_{mn} - k_{m}
\ell_{n} - k_{n} \ell_{m} \right), \quad k \cdot \ell = 1\,, \quad \ell
\cdot \ell = 0\,.
\end{equation}

\noindent
In the above equations, $\zeta_{mn}(x) = \zeta_{mn}(k) e^{i k\cdot x}$ and $k^m$ 
is the spacetime momentum. In order to describe spacetime
fermions, one should use either the vertex operator

\begin{equation}
\lambda A \widetilde{\lambda} = \left( \lambda \gamma^{m} \theta \right)
\left( \Ups_{n} (x) \gamma_{m} \theta \right) \left( \widetilde{\theta}
\gamma^{n} \widetilde{\lambda} \right),
\end{equation}

\noindent
or the vertex operator

\begin{equation}
\lambda A \widetilde{\lambda} = \left( \lambda \gamma^{m} \theta \right)
\left( \theta \gamma_{n} \widetilde{\Ups}_{m} (x) \right) \left( 
\widetilde{\theta} \gamma^{n} \widetilde{\lambda} \right),
\end{equation}

\noindent
where the spacetime fields $\Ups^{\alpha}_{m}$ and
$\widetilde{\Ups}^{\widetilde{\alpha}}_{m}$ describe the gravitini and
the dilatini. For RR fields one conventionally uses the vertex
operator

\begin{equation}
\lambda A \widetilde{\lambda} = \left( \lambda \gamma^{m} \theta \right)
\left( \theta \gamma_{m} F(x) \gamma_{n} \widetilde{\theta} \right)
\left( \widetilde{\theta} \gamma^{n} \widetilde{\lambda} \right),
\end{equation}

\noindent
where $F$ is the RR field strength,

\begin{equation}
F = \sum_{n}  \frac{1}{n!}  \gamma^{m_{1}} \cdots \gamma^{m_{n}}
F_{m_{1} \cdots m_{n}} (x)\,.
\end{equation}

\noindent
Here $n$ is even in type IIA and odd in type IIB. The
vertex operator associated with a constant RR gauge field
$C$ has also been studied \cite{Cornalba:2002} (see also \cite{Grassi:2004b}), 
and is

\begin{equation} \label{Cvertex}
\lambda A \widetilde{\lambda} = \left( \lambda \gamma^{m} \theta \right)
\left( \theta \gamma_{m} C \widetilde{\lambda} \right) - \left( 
\lambda C \gamma_{m} \widetilde{\theta} \right) \left( \widetilde{\theta}
\gamma^{m} \widetilde{\lambda} \right),
\end{equation}

\noindent
where

\begin{equation}
C = \sum_{n} \frac{1}{n!} \gamma^{m_{1}} \cdots \gamma^{m_{n}}
C_{m_{1} \cdots m_{n}}.
\end{equation}

\noindent
In this expression $n$ is even in type IIB and odd in type IIA. 
For later reference we note that in type IIB the vertex operator can 
also be written as 

\begin{equation} \label{UC}
\lambda A \widetilde{\lambda} =  - \left( 
\widetilde{\lambda} \widetilde{C} \gamma_{m} \theta \right) \left( \theta
\gamma^{m} \lambda \right)  - \left( 
\lambda C \gamma_{m} 
\widetilde{\theta} \right) \left( \widetilde{\theta}
\gamma^{m} \widetilde{\lambda} \right),
\end{equation}

\noindent
where $\widetilde{C}_{\al}{}^{\beta} = C^{\beta}{}_{\al}$, or explicitly,  

\begin{equation} 
\widetilde{C}_{\al}{}^{\beta} 
= \sum_{n \; \mathrm{even}} \frac{ (-1)^{\frac{n}{2}} }{n!} 
(\gamma^{m_{1} \cdots m_{n}})_{\al}{}^{\beta}
C_{m_{1} \cdots m_{n}}.
\end{equation}

\noindent
The integrated vertex operators can also be constructed in the same way as 
for the open superstring. 
For instance, the integrated vertex operator (in momentum space) 
corresponding to the Neveu--Schwarz--Neveu--Schwarz (NSNS) fields is given by 

\be
V \propto \zeta_{mn}\,(\pa X^m + i\, k_r L^{rn})
(\bar{\pa}x^n + i\, k_s \widetilde{L}^{sm} ) 
e^{i k\cdot X} + \cdots.
\ee

\noindent
As before, $\zeta_{mn}$ equals $h_{mn}$, $B_{mn}$ or $\ep_{mn} \zeta$, and the dots refer to terms with additional $\theta$'s and additional powers of $k$.   
Above we only discussed the massless modes of the string; vertex operators for 
the first massive level have been studied in~\cite{Berkovits:2002b}.

To compute tree amplitudes one also needs to deal with the zero
modes of $\theta^{\alpha}$ and $\lambda^{\alpha}$. It was argued in
\cite{Berkovits:2001a} that the pure spinor analogue of the RNS operator

\begin{equation} \label{RNSsatop}
c \partial c \partial^{2} c\ e^{-2 \phi},
\end{equation}

\noindent
which saturates the zero modes in tree amplitudes
\cite{Friedan:1985} is 

\begin{equation} 
\left( \lambda
\gamma_{m} \theta \right) \left( \lambda \gamma_{n} \theta \right)
\left( \lambda \gamma_{p} \theta \right) \left( \theta \gamma^{mnp} \theta \right).
\end{equation}

\noindent
This is the unique element of ghost number three 
in the cohomology of the BRST operator (\ref{QBRST}) \cite{Berkovits:2000a}. 
Tree amplitudes are
then obtained via $n$--point correlation functions with three unintegrated
vertex operators and $n{-}3$ integrated vertex operators, such that the zero
modes of $\theta^{\alpha}$ and $\lambda^{\alpha}$ are saturated via the
correlator \cite{Berkovits:2000a}

\begin{equation} \label{covsat}
\langle \left( \lambda \gamma^{m} \theta \right) \left( \lambda
\gamma^{n} \theta \right) \left( \lambda \gamma^{p} \theta \right)
\left( \theta \gamma_{mnp} \theta \right) \rangle = \mathrm{const}.
\end{equation}

\noindent
It has been checked that this prescription leads to results which are 
in complete agreement with the ones obtained from RNS~\cite{Berkovits:2000b}. 
Nevertheless, there remain some puzzling aspects about the saturation rule (\ref{covsat}). Some of those puzzles were resolved in the recent paper~\cite{Berkovits:2004a} where a refinement of the saturation rule was proposed.    
In the following, we shall show that this refinement can be obtained naturally 
by starting from the RNS saturation operator (\ref{RNSsatop}).

\subsection{Berkovits' Tree Level Zero Mode Saturation Rule from RNS} 
\label{sMes}

There is a seeming discrepancy between the (tree level) saturation rule (\ref{covsat}) for the zero modes in the pure spinor superstring, which we shall schematically write as $\lb \la^{3} \tha^5 \rb \neq 0$, and the one obtained by analysing (using the standard methods in \cite{Friedan:1985}) the background 
charges of $\pa h$ and $\pa g$ in (\ref{covst}), (\ref{covst2}). In fact, from (\ref{extra1}) and (\ref{extra2}), it follows that $\lambda^\alpha$ has $\partial h$ charge $\frac12$ while $\theta^\alpha$ has $\partial g$ charge $-\frac14$. Together with the fact that both Coulomb gas fields have background charge $4$, this naturally leads to the schematic saturation rule $\lb \la^{-8} \tha^{16} \rb \neq 0$ cancelling both background charges, but in contradiction with the previous expression. 
Recently Berkovits~\cite{Berkovits:2004a} proposed a refined version  
of the original saturation rule which resolves this mismatch: to relate 
the two saturation rules one further needs to insert eleven operators $Y^I$ 
(defined in~\cite{Berkovits:2004a}) into the original definition, 
each of which carries $(\la,\tha)$ charge $(-1,+1)$. 
Importantly, the properties of these operators are such that they 
do not affect earlier considerations in the literature based on the 
original rule (\ref{covsat}) and thus, for most purposes, 
they can be ignored.  
We should also point out that in earlier work by 
Chesterman~\cite{Chesterman:2004} equivalent, 
but less explicit, results were also obtained. In his work the existence of 
the different saturation rules was explained at the level of 
cohomology as arising from the isomorphism of certain cohomologies. 
In this language the extra $Y^I$ insertions can be viewed 
as the map implementing the isomorphism.

In this subsection we show how the modified saturation rule given 
in~\cite{Berkovits:2004a} arises from the saturation rule in 
the RNS superstring, using the map relating the two 
formulations~\cite{Berkovits:2001a}. 
Throughout this section we shall 
not keep track of numerical factors as they are not important for 
our conclusions. We will also only write the expressions for the 
left--moving sector explicitly; the right--moving sector is 
completely analogous.
To begin with, recall the bosonisation formul\ae{} 
of the RNS ghost variables $(\beta,\!\ga,b,c)$:

\be \label{RNSbos}
\beta = \pa \xi e^{-\phi}, \quad \ga = \eta e^{\phi}, \quad
\xi = e^{\chi}, \quad 
\eta = e^{-\chi}, \quad
c = e^{\si}, \quad b = e^{-\si} \,,
\ee

\noindent
from which it follows that (see, \textit{e.g.}, \cite{Polchinski:1998})

\be
:bc:=-\pa\si\,, \quad :\xi \eta: = \pa \chi \,, \quad
\de(\ga)= e^{-\phi} \,, \quad \de(\beta)=e^{\phi} \,.
\ee

\noindent
The bosonisation of the RNS world--sheet fermion $\Psi^m$ is (here $a=1,\ldots,5$)

\be
\Psi^{a}\pm i\Psi^{a+5} = e^{\mp \tau^a} \,.
\ee

\noindent
Later we will use the notation $\psi^a = e^{- \tau^a}$ and $\psi_a =e^{ \tau^a}$
As usual \cite{Friedan:1985} one constructs the spin fields from 
$e^{\pm \tau^1 \pm \tau^2\pm \tau^3 \pm \tau^4 \pm \tau^5}$
where the 32 different possibilities  
span a 32--dimensional Dirac spinor, which decomposes 
into a Weyl spinor $S^\al$, and an anti--Weyl spinor $S_\al$. These in turn 
decompose into $(S^+,S^a,S_{ab})$ and $(S_+,S_a,S^{ab})$ under the 
$\U(5)$ subgroup (see  appendix \ref{u5} for further details). 
Here we will only need 

\be 
S^a = e^{-\tau^a +\sum_b \tau^b/2}\,, \quad  S_a = e^{\tau^a -\sum_b \tau^b/2}
\,, \quad S^+ = e^{-\sum_a \tau^a/2} 
\,, \quad S_+ = e^{\sum_a \tau^a/2}\,.
\ee

\noindent
In the RNS theory, the well known saturation rules for the zero modes are 
(the first expression is just the bosonised version of (\ref{RNSsatop}))

\bea \label{RNSsat}
\lb e^{3\si -2\phi}\rb \neq 0 && \mbox{(small Hilbert space)} \,, \non \\
\lb e^{3\si -2\phi + \chi}\rb \neq 0 && \mbox{(large Hilbert space)}\,.
\eea

\noindent
To relate these expressions to the pure spinor result we first make 
the change of variables from RNS to the $\U(5)$ variables introduced 
in~\cite{Berkovits:1999,Berkovits:2001a} by Berkovits.  
In terms of these variables, a $\U(5)$ subgroup of the 
(Wick--rotated) $\SO(10)$ super--Poincar\'e symmetry is 
manifest~\cite{Berkovits:1999}. 
The $\U(5)$ variables comprise\footnote{As shown in \cite{Berkovits:2001a} 
these $\U(5)$ variables are the same  as (a subset of) the 
$\U(5)$ variables discussed in section \ref{flat-back}. 
As in \cite{Berkovits:2001a} the missing quartet 
$(p^{ab},\tha_{ab},v^{ab},u_{ab})$ will be added later.} 
12 Grassmann--odd variables 

\be \label{u5odd}
\theta^a = e^{\phi/2} S^a \,, \quad   
\theta^+   = c \, \xi \, e^{-3\phi/2}S^+ \,, \quad 
p_a = e^{-\phi/2} S_a \,, \quad 
p_+ =  b\,\eta\,e^{3\phi/2} S_+ \,,
\ee

\noindent
as well as the two Grassmann--even ones

\be \label{u5even}
s = \si -{\ts \frac{3}{2}}\phi -\half \sum_{a=1}^{5} \tau^a \,, \qquad
t = -\chi +{\ts \frac{3}{2}}\phi +\half \sum_{a=1}^{5} \tau^a \,.
\ee

\noindent
In the $(s,t)$ sector, the OPE is as (\ref{stOPE}) 
and the energy--momentum tensor is

\be
T = \pa s \pa t + \pa^2 s\,.
\ee

\noindent
By redefining these variables according to 
\be \label{st}
s = \half(\chihat - \phihat) \,, \qquad t = \chihat + \phihat \,,
\ee

\noindent
one finds 

\be
T = \half \pa \chihat \pa \chihat + \half \pa^2 \chihat 
- \half \pa\phihat\pa \phihat - \half \pa^2 \phihat\,,
\ee

\noindent
which one recognises (see, \textit{e.g.}, \cite{Polchinski:1998}) as
a bosonised $\behat\gahat$--system with weight $\la'=1$ 
($\la = \frac{3}{2}$) and therefore $T=-\behat\pa\gahat$. 
Note that this is {\em not} the same as the usual RNS $\beta\ga$--system. 
In particular, the conformal weight is different. 
The complete energy--momentum tensor also includes
$T= -p_a \pa \tha^a - p_+ \pa \tha^+$, \textit{i.e.}, six $bc$--type systems, 
all of weight one. In the large Hilbert space 
(with respect to the $\behat\gahat$--system) 
the saturation rule becomes (using standard methods~\cite{Friedan:1985})

\be
\lb e^{\chihat-\phihat} 
\ep_{abcde}\tha^a\tha^b\tha^c\tha^d\tha^e \tha^{+} \rb \neq 0 \quad
\leftrightarrow \quad \lb \xihat \de(\gahat)  
\ep_{abcde}\tha^a\tha^b\tha^c\tha^d\tha^e \tha^{+}  \rb \neq 0\,.
\ee

\noindent
Translating to the $(s,t)$ variables this becomes

\be
\lb e^{2s} 
\ep_{abcde}\tha^a\tha^b\tha^c\tha^d\tha^e \tha^{+} \rb \neq 0 \,,
\ee

\noindent
which agrees with what one obtains by translating the 
RNS result (\ref{RNSsat}) in the large Hilbert space 
(with respect to the $\beta\ga$--system), using 
the change of variables in (\ref{u5odd}) and (\ref{u5even}).  
This was to be expected since  
as long as one includes all zero modes (\textit{i.e.}, one works in the large Hilbert space) the saturation rule should be the same no matter which variables are used. 
On the other hand, in the small Hilbert spaces the saturation rules 
will not agree in general since they are defined with respect to 
different $\beta\ga$--systems. In the small $\U(5)$ Hilbert space 
(with respect to the $\behat\gahat$--system) 
the saturation rule is

\be
\lb e^{-\phihat} \ep_{abcde}\tha^a\tha^b\tha^c\tha^d\tha^e \tha^{+}  
\rb \neq 0 \quad \leftrightarrow \quad \lb \de(\gahat)  
\ep_{abcde}\tha^a\tha^b\tha^c\tha^d\tha^e \tha^{+}  \rb \neq 0 \,,
\ee

\noindent
which does not agree with what one obtains by translating the 
RNS result (\ref{RNSsat}) in the small RNS Hilbert space 
(with respect to the $\beta\ga$--system), using 
the corresponding change of variables.

In the map between RNS and pure spinor 
formulations~\cite{Berkovits:2001a}, the first step is to move to the 
large RNS Hilbert space. One would thus expect that after the change 
of variables it is the large $\U(5)$ Hilbert space which is relevant. 
However, one is also free to move to the small $\U(5)$ Hilbert space, and 
we will argue below that this is required in order to obtain the $\SO(10)$ 
covariant result.

To proceed, we add \cite{Berkovits:2001a} the ``topological'' 
quartet $(p^{ab},\tha_{ab},v^{ab},u_{ab})$ (\textit{cf.} section \ref{flat-back}).
This is a sum of ten $bc$-- and ten $\beta\ga$--type systems 
all with weight one. 
The additional operator insertion needed to saturate the zero modes 
is therefore:

\bea
\prod_{ab=1}^{10} \tha_{ab}\, \de(u_{ab}) \,.
\eea

\noindent
Note that this operator is in the small Hilbert spaces (with respect to the 
$(v^{ab},u_{ab})$ systems). 
Combining the above results it follows that 
the saturation rule can be written as

\be \label{cansat}
\lb \ep_{abcde}\tha^a\tha^b\tha^c\tha^d\tha^e
[\tha^{+} \prod_{ab=1}^{10} \tha_{ab} ]
[\de(\gahat) \prod_{ab=1}^{10} \de(u_{ab}) ] \rb \neq 0  \,.
\ee

\noindent
From (\ref{st}) 
and the discussion below (\ref{u5action}) we have $\gahat = e^{-2s}$, 
whereas from (\ref{lapar}) we 
find $\la^+ = e^s = \gahat^{-1/2}$.   
Using the relation $\de(x) = f'(x) \de(f(x))$ we 
find\footnote{The expression $(\la^+)^3 \de(\la^+)$, which naively 
might look like it is zero, is meaningless without the measure. We discuss 
the measure below; in the operator formalism the measure is implicit.}
$\de(\gahat) 
\propto (\la^+)^3 \de(\la^+)$.  Using this result together 
with (\ref{lapar}) we finally find

\be \label{prelsat}
\lb (\la^+)^3\ep_{abcde}\tha^a\tha^b\tha^c\tha^d\tha^e
[\tha^{+} \prod_{ab=1}^{10} \tha_{ab} ]
[\de(\la^+) \prod_{ab=1}^{10} \de(\la_{ab}) ] \rb \neq 0  \,,
\ee

\noindent
which is equivalent to

\be \label{satint}
\lb (\la \ga^m \tha) (\la \ga^n \tha) (\la \ga^p \tha) (\tha \ga_{mnp} \tha) 
[\tha^{+} \prod_{ab=1}^{10} \tha_{ab} ]
[\de(\la^+) \prod_{ab=1}^{10} \de(\la_{ab}) ] \rb \neq 0  \,,
\ee

\noindent
because the extra terms are set to zero by  
the delta functions (recall that for Grassmann--odd variables, 
$\de(\tha)=\tha$). One can write (\ref{satint})  as 

\be \label{endsat}
\lb (\la \ga^m \tha) (\la \ga^n \tha) (\la \ga^p \tha) (\tha \ga_{mnp} \tha) 
\prod_{I=1}^{11} C^I_{\al} \tha^{\al} \de(C^I_{\beta} \la^{\beta}) ] 
\rb \neq 0  \,,
\ee

\noindent
where the $C^I_{\al}$ are non--covariant constant spinors 
implicitly defined by the above two equations. The important point now is that  
$ C^I_{\al} \tha^{\al} \de(C^I_{\beta} \la^{\beta})$ is precisely what Berkovits 
called $Y_{C^I}$ in~\cite{Berkovits:2004a}, so that we have recovered his result in our setting. Above, our $C^I$'s were of a very special form, but 
it was shown in~\cite{Berkovits:2004a} that the expression (\ref{endsat}) is actually 
independent of the $C^I$'s. This completes our discussion of the 
relation between the RNS and pure spinor saturation rules. 

In the previous discussion we have exclusively worked within the operator formalism. 
Let us also briefly discuss the connection to the equivalent path integral 
approach and its associated zero mode measure. 
From the above expressions it might naively seem that 
$(\la^+)^3 \de(\la^+)$ is zero. But this expression is 
meaningless without the measure. 
The measure associated with (\ref{cansat}) is the canonical one 

\be \label{canmes}
\D \gahat \prod_{ab} \D u_{ab} \prod_{\al=1}^{16} \D\tha^{\al} \,.
\ee 

\noindent
Under the change of variables from $\gahat$ to $\la^+= \gahat^{-1/2}$ the measure 
$\D \gahat$ transforms into $\D \la^+ \frac{\D \gahat}{\D \la^{+}}  
\propto \D\la^+ (\la^+)^{-3}$, so that what remains after multiplication 
is $\int \D \la^+  \de(\la^+)$. After implementing this change 
of variables in (\ref{canmes}) we obtain 
the measure appropriate to the saturation rule (\ref{prelsat}). The form of 
the measure appropriate to the equivalent covariantised saturation rule 
(\ref{satint}) was constructed in \cite{Berkovits:2004a}. Note also that in the formulation where the pure spinor constraint is relaxed, some aspects of the measure were discussed in \cite{Grassi:2004}.

Let us end this subsection with a couple of comments. 
Notice that with our definitions it is not true that in the ghost sector 
$T = w_\al \pa \la^\al = w_+ \pa \la^+ + \half w^{ab} \pa \la_{ab}$, 
as can be seen by using (\ref{w}) and (\ref{lapar}). However, if one redefines 
$\la_{ab}\rar e^{c_1 s} u_{ab}$ and 
$w^{ab} \rar e^{-c_1 s} v^{ab}$, as well as $w_+ \rar e^{-s}(\pa t + c_2 \pa s 
+ c_3\, v^{ab} u_{ab})$, it is possible to choose the constants 
$c_1$, $c_2$ and $c_3$ in such a way that this is true. 
Also notice that one can define the analogue of the usual RNS picture number 
for the eleven $\beta\ga$--type systems: $(\behat,\gahat)$, 
$(v^{ab},u_{ab})$. In terms of these, the $Y^{I}$'s have picture number $-1$ 
(and ghost number $+1$).
Another point is that it can be checked that the operator saturating the zero modes is BRST closed. In the form (\ref{prelsat}) this is only true for the expression as a whole. On the other hand, for the covariantised expression (\ref{satint}) this is 
essentially true for the $\la^3 \tha^5$ and $\prod_I Y^I$ parts 
separately \cite{Berkovits:2004a},  
as required for the interpretation of the insertion of $\prod_I Y^I$ 
as a map between cohomologies \cite{Chesterman:2004}. 

The main result of the recent paper \cite{Berkovits:2004a} was a 
prescription for calculating loop amplitudes in the pure spinor 
formulation, which had been a long standing problem. 
In the calculation of loop amplitudes in the RNS formulation, insertions 
roughly of the form $\{Q,\Theta(\beta)\}$ occur, 
where $\Theta$ is the step--function. 
If one naively takes this to hold also in the pure spinor formulation, then 
one obtains (using (\ref{lapar}) and $\Theta'(x) = \de(x)$)

\be
\{Q,\Theta(v^{ab})\} = \half [d^{ab} e^s + {\ts \frac{1}{2}} 
\ep^{abcde}d_c u_{de} ] e^{-s}\, \delta(v^{ab}) 
 \propto (d \ga^{ab} \la) \, \delta( N^{ab}) \quad\mbox{(no sum)}
\ee

\noindent
where in the last step we have used (\ref{wlid1}). Introducing 
$B^{I}_{cd} = \de^a_c\de_d^b$, where $I=ab$, we can rewrite the 
above expression as 
$B^I_{cd} (d \ga^{cd} \la) \, \delta(B^I_{ef}N^{ef})$. Thus 
we find exactly the $Z_{B^I}$ 
operator defined in \cite{Berkovits:2004a}. 
Similarly, using $\behat \propto e^{2s}(\pa t + 2 \pa s)$, one finds

\be
\{Q,\Theta(\behat)\} \propto [e^s d^{+} - \la^a d_a] e^{2s}\, 
\delta(e^{2s} (\pa t + 2 \pa s )) 
\ee

\noindent
which is equal to $Q \,  \delta( J )$ modulo normal ordering 
ambiguities and terms which vanish when 
multiplied by $\de(u_{ab})$. Thus we find the operator $Z_J$ 
defined in \cite{Berkovits:2004a}. As we said earlier, it may be possible to make these arguments more precise. Also, related aspects of 
the loop amplitude prescription, as well as the geometry of the 
picture changing operators, were recently discussed in 
\cite{Grassi:2004c}.

\subsection{Mode Expansions}

In this subsection we collect the mode expansions of the world--sheet fields. 
Since this is a standard procedure, we shall be rather schematic. 
The world--sheet bosons $\partial x^{m}$, $\bar{\partial} x^{m}$, which 
transform as spacetime vectors, have the standard expansions

\begin{equation}
\partial x^{m} (z) = - i \sum_{k \in \mathbb{Z}}
\frac{\alpha^{m}_{k}}{z^{k+1}}\,, \qquad
\bar{\partial} x^{m} (\bar{z}) = - i \sum_{k \in
\mathbb{Z}} \frac{\widetilde{\alpha}^{m}_{k}}{\bar{z}^{k+1}}\,,
\end{equation}

\noindent
where the spacetime momentum is $p^{m} = \alpha^{m}_{0} = \widetilde{\alpha}^{m}_{0}$, and the commutation relations are
as usual $\left[ \alpha^{m}_{k}, \alpha^{n}_{l} \right] = k \, \eta^{mn}
\delta_{k+l}$ and similarly for $\widetilde{\al}^m_{k}$.
The world--sheet scalars which transform as spacetime fermions have integral 
world--sheet conformal dimensions, with mode expansions

\be
\theta^{\alpha} (z) = \sum_{k \in \mathbb{Z}}
\frac{\theta^{\alpha}_{k}}{z^{k}}\,, \qquad
p_{\alpha} (z) = \sum_{k \in \mathbb{Z}} \frac{p_{\alpha, k}}{z^{k+1}}\,, 
\ee

\noindent
and similarly for the right--moving fields. The commutation relations are, as
expected, $\{ p_{\alpha, k}, \theta^{\beta}_{l} \} =
\delta_{\alpha}^{\beta} \delta_{k+l}$ and similarly for 
the right--moving sector. The mode expansions of $\pa g$ and $M^{mn}$ are 
(we suppress the right--moving sector)

\begin{equation} \label{gMmodes}
\pa g (z) = \sum_{k\in \mathbb{Z} } 
\frac{g_k}{z^{k+1}}\,, \qquad 
M^{mn}(z) = \sum_{k\in \mathbb{Z} } 
\frac{M^{mn}_k}{z^{k+1}}\, .
\end{equation}

\noindent
As to the ghost field, $\la^\al$, its expansion is

\begin{equation}
\lambda^{\alpha} (z) = \sum_{k \in \mathbb{Z}} 
\frac{\lambda^{\alpha}_{k}}{z^{k}}\,. 
\end{equation}

\noindent
Finally, the expansions for $\pa h$ and $N^{mn}$ are completely 
analogous to the ones in (\ref{gMmodes}), and we shall not explicitly write them down.

The commutation relations involving the modes of 
$(M^{mn}, N^{mn}, \pa g, \pa h, \la^\al, \tha^\al)$ can easily 
be worked out; we have the following non--vanishing commutators 

\begin{eqnarray}
\left[ M^{rs}_{k}, M^{mn}_{l} \right] \!\!&=& \!\!\eta^{r m}
M^{s n}_{k+l} {-} \eta^{s m} M^{r n}_{k+l} {+} \eta^{r n} M^{s m}_{k+l} 
{-} \eta^{s m} M^{r n}_{k+l} {+} 4 l
\left( \eta^{m s} \eta^{n r} {-} \eta^{m r} \eta^{n s}
\right) \delta_{k+l}, \non \\
\left[ N^{rs}_{k}, N^{mn}_{l} \right] \!\!&=& \!\! \eta^{r m}
N^{s n}_{k+l} {-} \eta^{s m} N^{r n}_{k+l} {+} \eta^{r n} N^{s m}_{k+l} 
{-} \eta^{s m} N^{r n}_{k+l} {-} 3 l
\left( \eta^{m s} \eta^{n r} {-} \eta^{m r} \eta^{n s}
\right) \delta_{k+l}\,, \non \\
\left[ N^{mn}_{k}, \lambda^{\alpha}_{l} \right] \!\!&=& \!\!
\half {\left(\gamma^{mn} \right)^{\alpha}}_{\beta} 
\lambda^{\beta}_{k+l}\,,
\qquad \left[ M^{mn}_{k}, \theta^{\alpha}_{l} \right] = \half {\left(
\gamma^{mn} \right)^{\alpha}}_{\beta} \theta^{\beta}_{k+l} \,,\\
\left[ h_{k}, \lambda^{\alpha}_{l} \right] \!\!&=& \!\! \half
\lambda^{\alpha}_{k+l} \,, \qquad 
\left[ g_{k}, \tha^{\alpha}_{l} \right] = -{\ts \frac{1}{4} }
\tha^{\alpha}_{k+l} \,,\non \\
\left[ h_{k}, h_{l} \right] \!\!&=&\!\! -k \delta_{k+l}\,, \qquad 
 \!\!\!
 \left[ g_{k}, g_{l} \right] \, = \, k \delta_{k+l}\,. \non
\end{eqnarray}

\noindent
Finally, we should also point out that after a conformal transformation to the 
cylinder, the mode expansions are (in the $(\tau,\si)$ cylinder 
coordinates)  

\be \label{cylmodes}
F = \sum_k f_k e^{-i k (\tau - \si) } \,, \qquad 
\widetilde{F} = \sum_k \widetilde{f}_k e^{- i k (\tau + \si)}
\ee

\noindent
for any fields $F$, $\widetilde{F}$, irrespective of their conformal weights.

\subsection{Vacua}

We now discuss the vacuum structure of the theory in more detail. 
Let us first consider a simple toy model comprising one $bc$--system 
with fields $p$ and $\tha$ and one $\beta\gamma$--system 
with fields $v$ and $u$. We assume that both systems have conformal weight equal to one. For general weights, one needs to distinguish between the oscillator vacuum and the $\mathrm{SL}(2)$ vacuum. However, for the special case of weight one, these 
coincide. We denote the $\mathrm{SL}(2)$ invariant oscillator vacuum 
by $\vack$. This state satisfies $L_0\vack =0 = L_{\pm}\vack$. 
In terms of the oscillator modes in the ``$bc$'' sector this state satisfies

\be
 p_k \vack = 0  \ \ \mbox{for}\ k \geq 0 
\quad \mathrm{and} \quad 
\theta_k \vack = 0 \ \ \mbox{for}\ k \geq 1\,. 
\ee

\noindent
As is well known there is another 
state with the same energy, namely, $\tha_0 \vack$. This state has ``ghost'' number 
$+1$. The notation $| \! \uparrow \,\rb$ is often used for this state whereas 
the notation $| \! \downarrow \,\rb$ is often used for what we previously called $\vack$. The only non--vanishing matrix element between the degenerate states 
is $\vacb \tha_0 \vack$ (or using the other notation, 
$\lb \, \downarrow  |  \uparrow \, \rb$). 
In the ``$\beta\gamma$'' sector, $\vack$ satisfies

\be
  v_k \vack = 0 \ \ \mbox{for}\ k \geq 0 
\quad \mathrm{and} \quad 
  u_k \vack = 0 \ \ \mbox{for}\ k \geq 1\,. 
\ee

\noindent
As was the case for the $bc$--system, there are other vacuum states. 
In particular, the state which has a non--vanishing matrix element 
with $\vacb$ is $\de(u_0)\vack$ (this is usually shown using bosonisation). 
The state $\de(u_0)\vack$  has ``picture'' number $-1$.
To conclude, there are two natural vacuum states one can define, $\vack$ and 
$|\Om\rb = \tha_0 \, \de(u_0) \vack$, and these states have a non--vanishing 
inner product as $\vacb \Om\rb \neq 0$.

Let us now make contact with the saturation rule given in (\ref{satint}). 
This rule can be written as (here we suppress the `$0$' 
zero mode subscripts)

\be \label{satvac}
\vacb (\la \ga^m \tha) (\la \ga^n \tha) (\la \ga^p \tha) (\tha \ga_{mnp} \tha) 
| \Om \rb \neq 0\,,
\ee

\noindent
where 

\be
|\Om \rb = \prod_{I=1}^{11} Y^I \vack = [\tha^{+} \de(\la^+)]
\prod_{ab=1}^{10}[ \tha_{ab} \de(\la_{ab}) ] \vack\,.
\ee 

\noindent
Recall that there are eleven $(b,c,\beta,\gamma)$--type quartets 
of weight one in the $\U(5)$ formulation of the pure spinor superstring:  
$(p_+,\tha^+,\behat,\gahat)$ and the ten $(p^{ab},\tha_{ab},v^{ab},u_{ab})$. 
Given the above discussion we see that the $Y^I$ insertions can essentially 
be understood as a change in vacuum, required to get a non--vanishing 
matrix element (since $\de(\la^+)$ is used rather than $\de(\gahat)$ 
there is a remaining background charge which is cancelled 
by the explicit additional $\la^3$ insertion).

\setcounter{equation}{0}
\section{Pure Spinor Boundary States in Flat Spacetime} \label{secbs}

In this section we discuss the construction of the D--brane boundary state, in 
the pure spinor superstring. Throughout we restrict ourselves to 
the case of a flat spacetime background. Some of the results in 
subsections~\ref{openpic} through \ref{bnzm}, and \ref{lowerD} were also obtained in \cite{Anguelova:2003}, albeit using a slightly different approach. However, the zero mode part of the boundary state was not discussed in~ \cite{Anguelova:2003}. 
For simplicity we shall concentrate on the D$9$--brane in the type IIB theory (but a brief discussion of the lower dimensional D$p$--branes can be found at the end of this section).

\subsection{The Open String Picture: Sigma Model Boundary Conditions}
\label{openpic}

We start by considering an open string stretched between two
D$9$--branes in the type IIB string theory. 
The relation between the $(z,\bar{z})$ coordinates and 
the $(\tau,\sigma)$ coordinates is

\begin{equation}
z = e^{i(\tau-\sigma)},
\quad
\bar{z} =  e^{i(\tau+\sigma)},
\qquad 
\partial_\tau = i \left( z \partial + \bar{z} \bar{\partial} \right),
\quad
\partial_\sigma = i \left( - z \partial + \bar{z} \bar{\partial} \right).
\end{equation}

\noindent
We are considering an open string with endpoints at $\sigma=0$ and
$\sigma=\pi$. To determine the 
boundary conditions that the open string should obey at its endpoints we 
consider the variation of the action (\ref{flat-action}). At this point 
we do not consider the ghost sector; this sector will be included later. 
The  variation of the action gives the equations of
motion for the fields, plus the boundary term

\begin{equation}
\frac{1}{2\pi} \int d\tau \left[ -2 \, i \left(z \partial - \bar{z}
\bar{\partial} \right) 
x_m \delta x^m -2 \, i \, z \, p_\alpha \delta \theta^\alpha 
+ 2 \, i \, \bar{z} \, \widetilde{p}_{\al} \delta 
\ttha^{\al}
\right]_{\sigma=0}^{\si =\pi} \,,
\end{equation}

\noindent
that needs to be set to zero. Consider $\sigma=0$ corresponding 
to $z= \bar{z}$. The boundary condition on $x^m$ is

\begin{equation}
\delta x^m \left(z \partial - \bar{z} \bpart \right) 
x_m \Big|_{\sigma=0}=0 \,.
\end{equation}

\noindent
This is the usual condition on the bosonic modes of the open string. 
For a D$9$--brane we impose Neumann boundary conditions $\left( 
z\partial - \bar{z} \bar{\partial} \right) x^m |_{\sigma=0} = 0$. In terms 
of the $(\tau,\si)$ coordinates this is just $\pa_{\si} x^{m} |_{\sigma=0} = 0$.

Let us now consider the spacetime fermionic sector. It is clear that one needs

\begin{equation}
\label{pthcond}
\left( z \, p_\alpha \delta \theta^\alpha - \bar{z} \,
\widetilde{p}_\al \delta \ttha^\al \right)
\Big|_{\sigma=0} = 0\,.
\end{equation}

\noindent
The boundary term thus 
vanishes if one imposes the boundary conditions 

\begin{equation}
\label{pthcond2}
\left( z \, p_\alpha  \pm  \bar{z} \, \widetilde{p}_\al \right)
\Big|_{\sigma=0} = 0,
\quad
\left( \theta^\alpha \pm  \ttha^\al \right)
\Big|_{\sigma=0} = 0 \,.
\end{equation}

\noindent
From the above results (written in terms of the oscillator modes) 
it follows that the boundary conditions on 
the supersymmetry generators (\ref{SuzyQ}) are\footnote{Observe that in (\ref{qcond1}) we are actually referring to the supersymmetry current and not to the supersymmetry charge. We will return to the standard notation in what follows.}

\begin{equation}
\label{qcond1}
\left( z \, q_\alpha  \pm
\bar{z} \, \widetilde{q}_{\al} \right) \Big|_{\sigma=0} = 0 \,.
\end{equation}

\noindent
This implies the well--known fact that the D--brane 
preserves half the supersymmetries, 
and justifies the conditions (\ref{pthcond2}).
It also follows from the above conditions that the fermion Lorentz currents
$M^{m n} = - \frac{1}{2} p \gamma^{m n} \theta$, $\widetilde{M}^{m n} 
= - \frac{1}{2} \widetilde{p} \gamma^{m n} \ttha$, and the fermion Lorentz scalars 
$g= \frac{1}{4}\, p \theta$, $\widetilde{g}= \frac{1}{4} \, \widetilde{p}  \ttha$, 
obey the boundary conditions 

\begin{equation}
\label{ferlor}
\left( z M^{m n} - 
\bar{z} \widetilde{M}^{m n} \right) \Big|_{\sigma=0} = 0 \,, \qquad 
\left( z \pa g - \bar{z} \bar{\pa} \widetilde{g} \right) \Big|_{\sigma=0} = 0  \,.
\end{equation}

\noindent
One should not confuse the $\pm$--ambiguity above with the one appearing in the RNS superstring. Whereas in the RNS case this ambiguity is related to the world--sheet spin structures, the ambiguity here should rather be regarded as a spacetime ambiguity associated to a choice of brane versus anti--brane. That this is so can be understood by the fact that a D--brane breaks half of the target space supersymmetries and, as one can see from (\ref{qcond1}) above, there are two distinct ways to do so: one must be associated to a D--brane and the other 
to the anti--D--brane. Later, we will indeed see that this sign ambiguity corresponds to objects with the same tension but opposite RR charges, as expected.

\subsection{The Closed String Picture: D--Brane Boundary States}

If one computes the open string one--loop amplitude, with the boundary
conditions just found in section \ref{openpic}, for open strings ending on a
D--brane, one is computing the annulus diagram for the open
string. This annulus diagram can equivalently be described in terms of  
closed string propagation, as is well--known. 
In order to go from the open string picture to the closed
string picture, one essentially needs to do the transformation $\tau 
\leftrightarrow \si$ 
(see, \textit{e.g.}, \cite{DiVecchia:1999b} for a detailed discussion). 
In general one also has the option of writing the relevant expressions 
either in terms of the $(z,\bar{z})$ coordinates on the complex plane, or to make 
a conformal transformation to the $(\tau,\si)$ cylinder coordinates. 
Under the latter transformation the fields will have 
fixed transformation properties 
dictated by their conformal weights, \textit{e.g.}, $\tha^\al \rar \tha^\al$ and 
$\ttha^\al\rar \ttha^\al$, whereas $z p_\al \rar i p_\al$ and
$\bar{z} \widetilde{p}_\al \rar -i \widetilde{p}_\al$.

Via the above transformations, one can directly translate the
boundary conditions for the open string, obtained in section \ref{openpic}, 
into conditions for the (closed string sector) boundary state $\Bk$. 
Below, all conditions are written in terms of the $(\tau,\si)$ variables, and 
all fields have mode expansions as in (\ref{cylmodes}). We find

\begin{eqnarray}
& 
\left. \partial_{\tau} x^m
\right|_{\tau=0} | B \rangle = 0\,, & \\
& 
\left. \left( p_\alpha \mp  \widetilde{p}_\al \right)
\right|_{\tau=0} | B \rangle = 0\,, 
\quad
\left. \left( \theta^\alpha \pm  \ttha^\al 
\right) \right|_{\tau=0} | B \rangle = 0\,, & \\
&
\left. \left( M^{m n} +  \widetilde{M}^{m n} \right) 
\right|_{\tau=0} | B \rangle = 0 \,,  \quad 
\left. \pa_{\tau}( g + \widetilde{g} ) 
\right|_{\tau=0} \Bk \,.&
\end{eqnarray}

\noindent
In terms of the modes of the fields this translates into

\begin{eqnarray}
&
\left( \alpha^m_k + \widetilde{\alpha}^m_{-k} \right) | B \rangle = 0 \,, &
 \\
&
\left( p_{\alpha,k} \mp \widetilde{p}_{\al,-k} \right) | B \rangle = 0\,, 
\qquad 
\left( \theta_k^\alpha \pm \ttha^\al_{-k} \right)  | B \rangle = 0\,, & \\
&
\left( M_k^{m n} + \widetilde{M}_{-k}^{m n} \right) | B \rangle = 0 \,,
\qquad
\left( g_k + \widetilde{g}_{-k} \right) | B \rangle = 0 
\,. &
\end{eqnarray}

\noindent
for any $k \in \mathbb{Z}$.

We now turn our attention to the ghost sector. It is clear that one could 
repeat the above analysis for the free $\U(5)$ ghost variables. However, rather than giving the details of such an analysis, we shall instead present another 
approach in what follows (which leads to the same results).
First, notice that the boundary state should satisfy 

\be
\left( Q + \widetilde{Q} \right) \Bk = 0 \,,
\ee

\noindent
where the BRST operators were given in (\ref{QBRST}). Consistency 
of this relation with the boundary 
condition on $(\tha^\al,\ttha^\al)$ implies that 

\be
\left. \left( \la^\al  \pm  \widetilde{\la}^\al \right) \right|_{\tau=0} 
\Bk = 0\,,
\ee

\noindent
which in turn is consistent with the conditions

\begin{equation}
\label{ghostlor}
\left. \left( N^{m n} + \widetilde{N}^{m n} \right) \right|_{\tau=0} \Bk = 0 
\,, \quad
\left.  \pa_{\tau} (h + \widetilde{h})  \right|_{\tau=0} 
\Bk = 0 \,.
\end{equation}

\noindent
A further consistency check is that the Lorentz generators in 
the spacetime fermionic sector and in the ghost sector satisfy the same 
boundary conditions, as required. Having obtained the boundary conditions, we 
next turn to the construction of the state $\Bk$ 
satisfying the above conditions.

\subsection{Boundary State: Non--Zero Mode Sector}
\label{bnzm}

Without loss of generality one can write $\Bk$ as $e^W\Bk_0$ where 
the prefactor $e^W$ only involves non--zero modes, and $\Bk_0$ 
refers to the zero mode part. The part of the $e^W$ prefactor involving 
the bosonic modes of $x^m$  is as usual

\begin{equation}
\exp \left( - \sum_{k=1}^\infty \frac{1}{k} 
\alpha_{m,-k} \, \widetilde{\alpha}_{-k}^m \right) .
\end{equation}

\noindent
Similarly, the corresponding $p\theta$ prefactor is 

\begin{equation}
\exp \left( \mp \sum_{k=1}^\infty \left[ p_{\alpha,-k} \, \ttha^\al_{-k} 
+ \widetilde{p}_{\al,-k}\, \theta_{-k}^\alpha \right] \right) .
\end{equation}

\noindent
The ghost part of the prefactor can, \textit{e.g.}, be 
written in terms in terms of the $w\la$ variables as (we could 
also have written it in terms of the free $(\behat,\gahat,v^{ab},u_{ab})$ 
variables and their right--moving cousins)

\be \label{wlnzm}
\exp \left( \pm \sum_{k=1}^\infty  \left[
w_{\al,-k} \, \widetilde{\la}_{-k}^\al + \widetilde{w}_{\al,-k} \, \la_{-k}^\al 
\right] \right) .
\ee

\noindent
It is important to observe that although this expression is seemingly not manifestly covariant (because of the presence of $w_\al$, $\widetilde{w}_{\al}$), 
the boundary conditions 
it implies are indeed covariant. Notice that when (\ref{wlnzm}) is written 
in $\U(5)$ components, 
only the canonically conjugate variables $(w_+,\la^+)$, $(w^{ab},\la_{ab})$ 
(and their right--moving counterparts) appear,    
since $w_a=0=\widetilde{w}_a$ (\textit{cf.} appendix \ref{u5}). It is easy to see that this is not in conflict with the boundary condition 
$(\la^a \pm \tla^a)| B \rangle=0$ because of the composite nature of 
$\la^a$, $\tla^a$ implied by the pure spinor constraint.
Let us also mention that the construction of the cross--cap state $|C\rb$ 
is completely analogous, the only difference being that one finds 
extra $(-1)^k$ insertions in the sums in the above expressions. 

In this paper the above explicit expressions for the non--zero mode part 
of the boundary state will play a marginal role, 
so we shall not discuss them any further in this work. Instead we turn 
to the much harder problem of determining $\Bk_0$---the zero mode 
part of the boundary state.

\subsection{Boundary State: Zero Mode Sector}

As discussed above, the boundary state $\Bk$ contains two parts: 
a part containing the non--zero modes and a part containing 
only zero modes. 
There are several different possible routes one could take to 
obtain the zero mode part of the covariant boundary state in 
the pure spinor superstring. Some possibilities are:

\begin{itemize}
\item Start from the known expression in the RNS superstring and try to 
apply the map in \cite{Berkovits:2001a} in order to obtain a result in terms 
of the pure spinor variables. 
\item Start from RNS, written in its $\U(5)$ version \cite{Berkovits:1999}, and derive the boundary states 
directly in terms of these variables, then translate the result so obtained into 
the pure spinor superstring.
\item Start directly with the pure spinor variables, alongside with the associated boundary conditions, and try to derive the boundary state.
\item Use known information about the pure spinor massless vertex 
operators (the zero mode cohomology) in order to try 
to obtain information about the zero mode part of the boundary state. 
\end{itemize}

\noindent
Below we will comment on each of these approaches. As above, we 
restrict ourselves to the D$9$--brane in type IIB; the more general case will be 
discussed later. Also, in order not to clutter the formul\ae{}, we 
shall not explicitly write the index `$0$' on the various variables 
to indicate that, in this subsection, we are dealing exclusively
with the zero modes. From the previous discussion it follows that 
we want the zero mode part of the boundary state to satisfy 
the conditions

\be \ba{rrr} \label{psbcB}
(\tha^\al \pm \ttha^\al) \Bk_0 =0\,, 
& (p_\al \mp \tp_\al) \Bk_0=0 \,,
& (\la^\al \pm \tla^\al) \Bk_0=0 \,, \\
 (Q + \widetilde{Q}) \Bk_0=0 \,, 
& (L^{mn} + \widetilde{L}^{mn}) \Bk_0=0 \,,
& (q_{\al} \mp \widetilde{q}_{\al})\Bk_0=0\,.
\ea
\ee

\noindent
The first condition, together with the last, implies 
that the $\tha,\ttha$ dependence of the zero mode part of $\Bk$ has to be  
$\Bk_0^{\tha,\ttha}=\prod_{\al=1}^{16} (\tha^\al \pm \ttha^\al) |0\rb$, 
where $|0\rb$ is annihilated by $p_\al$ and $\tp_{\al}$ (additional factors 
of $\tha,\ttha$ are ruled out by the fact that we want precisely half the 
supersymmetry generators to annihilate the boundary state).

In the ghost sector the situation is more involved. 
It is clear that we want a factor $\de(\la^\al \pm \tla^\al)$ for each 
independent component of $\la,\tla$. However, since the $\la,\tla$ 
variables are bosonic an additional multiplicative function of $\la,\tla$ 
might also be allowed and it is not immediately obvious 
how to choose this function. 
The approach we shall now take to determine the ghost part of the boundary state is based on the free $\U(5)$ variables; at the end we will check that 
all boundary conditions are satisfied.

In terms of the $\U(5)$ ghost variables (which we recall comprise 
eleven $\beta\ga$--systems, all of conformal weight one) the natural candidate 
for the zero mode part of the boundary state 
is $\Bk_0^{\gahat, u;\gathat,\widetilde{u}} = \de(\gahat - \gathat)
\prod_{ab} \de(u_{ab}  \pm \widetilde{u}_{ab})|0\rb$, where $|0 \rb $ has 
picture number 0 and is 
annihilated by $\behat$, $\bethat$, $v^{ab}$ and 
$\widetilde{v}^{ab}$. Strictly speaking there are some sign ambiguities here.  
We have chosen to fix the signs so that Lorentz covariance is recovered at 
the end (see the later discussion). Recalling that $\gahat = (\la^+)^{-2}$ 
(and similarly for $\gathat$), we see that $\gahat = \gathat$ 
implies $\la^{+}=\mp\tla^+$. The transformation of  
$\de(\ga - \gahat)$ to the $\la^+$, $\tla^+$ variables gives 
$\de(\ga - \gahat) \propto 
(\la^+)^{k} (\tla^+)^{3-k} \de(\la^+ \pm \tla^+)$, 
where the integer $k$, with $0\le k\le 3$, is arbitrary. This arbitrariness 
arises from the presence of the delta function (there is a similar 
ambiguity in the radial part of the usual three dimensional delta function 
expressed in spherical coordinates). In 
the path integral approach the integration is performed using 
the closed string measures $\int \D \la^+ (\la^{+})^{-3}$ 
and $\int \D \tla^+ (\tla^{+})^{-3}$; this shows that all $k$ 
give identical results.

Combining the above results we (schematically) obtain

\be \label{B0prel}
\Bk_{0} \propto \sum_{n=0}^{5} (\pm)^n (\la^+)^{k_{n}} 
(\tla^+)^{3-k_{n}} \tha^n \ttha^{5-n} 
\prod_{I=1}^{11} Y^{I}(\tha \pm \ttha,\la \pm \tla)|0\rb \,
\ee

\noindent
where, as in section \ref{sMes}, we  have re--written the delta functions 
involving $\tha^+$, $\tha_{ab}$, $\la^+$,and $\la_{ab}$ 
in terms of 
$Y^I(\tha,\la) = C^I_{\al} \tha^{\al}  \, \de( C^I_{\al}\la^\al)$ 
for certain constant spinors $C^I_\al$. 
The remaining $\tha^a$, $\ttha^a$ variables 
have been written using a short--hand notation, 
\textit{e.g.}, $\tha^2 \ttha^{3} = \ep_{abcde}\tha^a \tha^b \ttha^c\ttha^d\ttha^e$ and so on. The basic boundary conditions for the free fields discussed above 
can be used to show that the boundary state (\ref{B0prel}) satisfies 
the correct boundary conditions also for the composite fields 
$Q$, $q_\al$, $M^{mn}$, $\pa g$, $N^{mn}$ and $\pa h$, by using their 
explicit expressions in terms of the free fields.   

We will now argue that, without changing the physics, one can replace the 
$(\la^+)^{k_{n}} (\tla^+)^{3-k_{n}} \tha^n \ttha^{5-n}$ 
pieces in the boundary state 
by manifestly covariant expressions, 
for instance one may replace $\la^+ (\tla^+)^{2} \tha \ttha^{4} \rar 
(\la \ga^m \tha)(\tla \ga^n \ttha)(\tla \ga^p \ttha)(\ttha \ga_{mnp} \ttha)$. 
One direct argument is to note that the presence of the 
$\de(\la^\al-\tla^\al)$'s in the $Y^I$'s means that in 
the covariant expressions 
we can replace, \textit{e.g.}, all $\tla^\al$'s by $\la^\al$'s. 
The remaining $\theta$, $\ttha$ dependence is then via 
$\tha^\al-\ttha^\al$, but the fermionic delta functions in the $Y^I$'s 
then put the $^+$ and 
${}_{ab}$ components to zero. Finally, one notes that there is no way 
to form a non--vanishing scalar out of a product of five  
$(\tha-\ttha)^a$'s as well as $\ep_{abcde}$ and  $\la^+$, 
which {\it also} involves $\la^a$ and/or $\la_{ab}$. Thus 
only the $\la^+$ dependence survives and we are back to the above 
expression (\ref{B0prel}).

An indirect but perhaps more transparent argument is to note that 
in calculating correlation functions (and as we shall later see, when 
calculating scattering amplitudes) one also needs to insert 
$\prod_{I=1}^{11} Y^{I}(\tha \mp \ttha,\la \mp \tla)$ to saturate 
the zero modes. Using the fact that 

\be
\prod_{I=1}^{11} Y^{I}(\tha \mp \ttha,\la \mp \tla)
Y^{I}(\tha \pm \ttha,\la \pm \tla) 
\propto \prod_{I=1}^{11} Y^{I}(\tha,\la)Y^{I}(\ttha,\tla) 
\equiv \prod_{I=1}^{11} Y^{I} \widetilde{Y}^{I}
\ee

\noindent
one realises that the presence of 
the delta functions in the $Y^I$ and $\widetilde{Y}^I$'s means that 
in the covariant expressions only the terms  present in (\ref{B0prel}) survive.

There still remains the ambiguity in choosing the $k_n$'s, though. Although 
this ambiguity does not appear to have any direct physical meaning 
there is a certain choice which is quite convenient.  
To describe this choice we recall that the left-- and right--moving parts of the 
vertex operators for the massless states involve certain basic constituent
operators. These operators essentially coincide with the 
zero mode cohomology (which was explicitly 
worked out in \cite{Berkovits:2001b}; see 
also~\cite{Cederwall:2001a,Cederwall:2001b,Chesterman:2004}) 
and are summarised in the table below. 

\vspace*{2pt}
\begin{center}
\begin{tabular}{|c|c|}
\hline 
$(p,q)$ &  $U_{p,q}$ \\ 
\hline
$(0,0)$ & $1$ \\ 
$(1,0)*$ &  $\la^\al$ \\
$(1,1)$ &  $(\la \ga^m \tha)$ \\
$(1,2)$ &$ (\la \ga^m \tha)(\tha\ga_m)_{\al}$ \\
$(2,3)$ &  $(\la \ga^m \tha) (\la \ga^n \tha) (\ga_{mn}\tha)^{\al}$ \\
$(2,4)$ & $(\la \ga^m \tha) (\la \ga^n \tha) (\tha \ga_{mnp} \tha)$ \\
$(2,5)*$ & $(\ga^m \tha)_\al(\la \ga^n \tha) (\la \ga^p \tha) 
(\tha \ga_{mnp} \tha)$ \\
$(3,5)$ & $(\la \ga^m \tha) (\la \ga^n \tha)(\la \ga^p \tha) 
(\tha \ga_{mnp} \tha)$ \\
\hline
\end{tabular} 
\\[8pt]
{\footnotesize {\bf\sf Table 2:} Basic ``massless'' operators. 
The entries marked 
with * are not in the BRST cohomology.}
\\[5pt]
\end{center}

\noindent
These operators are normalised according to

\be
\langle U_{p,q} U_{r,s} \rangle = c \, \de_{3-p-r} \, \de_{5-q-s}\,,
\ee

\noindent
where $c$ is a constant which depends on conventions and which we shall take 
to be equal to 1. In the above formula we 
have suppressed the index contractions, as well as the extra $Y^I$ insertions 
needed to get a non--vanishing answer.
Guided by the properties of the $U_{p,q}$'s we fix the $k_n$ ambiguity 
so that the ``covariant'' part of the  zero mode boundary state 
contains \textit{only} operators of this type. By comparison 
with the situation in RNS we also require the terms in 
the ``covariant'' part of the zero mode part 
of the boundary state to have $(\la,\tla)$ ghost numbers $(1,2)$ or 
$(2,1)$. This choice will be further discussed below.
Our way of fixing the $k_n$ ambiguity implies that although 
$Q+\widetilde{Q}$ of course annihilates the full 
boundary state by construction, it now also almost annihilates 
the ``covariant'' part (\textit{i.e.}, the part without 
the $Y^I$'s) separately without using the delta functions. We say 
almost, because for the states with five 
$\tha$'s (five $\ttha$'s) we choose $U_{2,5}\widetilde{U}_{1,0}$ 
($U_{1,0}\widetilde{U}_{2,5}$) rather than $U_{3,5}$ ($\widetilde{U}_{3,5}$).

From these requirements it follows that the zero mode part of 
the boundary state takes the final form

\be \label{bs0}
\Bk_{0} = \sum_{i=0}^{2} \left[ (\pm)^i U_{2,i{+}3} 
\widetilde{U}_{1,2{-}i} + (\pm)^{i+1} 
U_{1,2{-}i} \widetilde{U}_{2,i{+}3} \right]
\prod_{I=1}^{11} Y^{I}(\tha \pm \ttha,\la \pm \tla)|0\rb \,,
\ee

\noindent
where $U_{p,q}$ ($\widetilde{U}_{p,q}$) is a state in table 2 
of the schematic form $\la^p \tha^q$ ($\tla^p\ttha^q$). 
In this expression indices have been 
suppressed (index contractions are the natural ones).

Equation (\ref{bs0}) is our final result for the zero mode part of the 
boundary state. It was constructed to satisfy 
the boundary conditions (\ref{psbcB}). 
From this requirement we showed that it could be written as a part containing 
$Y^I$'s, times another part which, by the arguments given above, 
could be written in a manifestly covariant manner and which we chose to express 
in terms of the operators in table 2. Suppressing the $Y^I$ insertions 
by defining  $|Z\rb =\prod_{I=1}^{11} Y^{I}(\tha \pm \ttha,\la \pm \tla)|0\rb$, 
the zero mode part of the boundary state takes the 
form 

\be \label{bs0cov}
\Bk_{0} = \sum_{i=0}^{2} \left[ (\pm)^i U_{2,i{+}3} 
\widetilde{U}_{1,2{-}i} + (\pm)^{i+1} 
U_{1,2{-}i} \widetilde{U}_{2,i{+}3} \right] |Z\rb\,,
\ee

\noindent
where $U$ ($\widetilde{U}$) are the right-- (left--) moving 
``open string'' vertex operators for the massless fields given in table 2. 
Such a form was expected on general grounds, as well as 
from comparison with RNS, where similar 
results hold \cite{Callan:1988} (see also the discussion below). 
The main rationale behind our choice for fixing the $k_n$ ambiguities is that 
this choice is such that, when performing calculations, the $Y^I,\widetilde{Y}^I$ 
insertions largely decouple and one can essentially ignore them and 
effectively use the above form (\ref{bs0cov}), together with the original 
saturation rule (\ref{covsat}). 
However, one must always keep in mind that the $Y^I$ insertions are crucial to 
make sure that $\Bk_0$ satisfies the correct boundary conditions. 
For other choices of fixing the ambiguity one would have to use the 
delta functions in the $Y^I$'s to replace $\la$'s by $\widetilde{\la}$'s 
at certain intermediate stages of the calculations. 
We should also point out that there 
may well be a more  physical way to understand our expression, perhaps starting from the observation that the covariant piece in the zero mode boundary state 
(\ref{bs0cov}) has total ghost charge three and total spacetime fermion charge five (by total we mean sum of left and right), as does the pure spinor saturation rule.

As a final comment, and in order to gather some additional circumstantial support 
for our proposal, we now briefly 
(and at a very heuristic level) discuss the relation to 
the RNS expressions. In the RR sector the zero mode part of 
the RNS boundary state involves expressions of the form 

\be
\ba{lll}
& e^{2\si} e^{\widetilde{\si}}  e^{-1/2\phi} e^{-3/2\widetilde{\phi}} 
S_\al \widetilde{S}^{\al} 
=  e^{2\si} e^{-1/2\phi} e^{\widetilde{\si}} e^{-3/2\widetilde{\phi}} 
S_+ \widetilde{S}^{+} +\cdots 
&\propto (\la^+)^2\widetilde{\la}^+  \tha^5 + \cdots \,, \non \\
& e^{\si} e^{2\widetilde{\si}}  e^{-1/2\phi} e^{-3/2\widetilde{\phi}} 
S_\al \widetilde{S}^{\al} 
= \half  e^{\si} e^{-1/2\phi} e^{\widetilde{2\si}} e^{-3/2\widetilde{\phi}} 
S^{ab} \widetilde{S}_{ab} +\cdots 
& \propto \la^+ (\widetilde{\la}^+)^2 \tha^2\ttha^3  +\cdots \,, \non \\
& e^{\si} e^{2\widetilde{\si}} e^{-3/2\phi} e^{-1/2\widetilde{\phi}} 
S^\al \widetilde{S}_{\al} = 
e^{\si} e^{-3/2\phi} e^{2\widetilde{\si}}  e^{-1/2\widetilde{\phi}} 
S^+ \widetilde{S}_{+} + \cdots 
& \propto \la^+ (\widetilde{\la}^+)^2 \ttha^5  + \cdots \,, \non \\
& e^{2\si} e^{\widetilde{\si}} e^{-3/2\phi} e^{-1/2\widetilde{\phi}} 
S^\al \widetilde{S}_{\al} = \half 
e^{2\si} e^{-3/2\phi} e^{\widetilde{\si}}  e^{-1/2\widetilde{\phi}} 
S_{ab} \widetilde{S}^{ab} +\cdots 
& \propto (\la^+)^2 \widetilde{\la}^+ \tha^3\ttha^2  +\cdots \,, \non
\ea
\ee

\noindent
where we used a short--hand notation, 
\textit{e.g.}, $\tha^2 \ttha^3 = \ep_{abcde}\tha^a\tha^b\ttha^c\ttha^d\ttha^e$.
Thus one can surmise the appearance of four of the six terms in (\ref{bs0}).

In the NSNS sector things are more subtle since there one has to distinguish 
between the zero mode sector, and the part of the boundary state 
which couples to the massless fields. In the pure spinor case there 
is no such distinction, so what one should try to map to 
the pure spinor zero mode boundary state (which is also the part 
which couples to the massless sector) is the part of 
the NSNS boundary state which couples to the massless modes, namely 

\be
\Bk_{m^2=0, \mathrm{NSNS}} \propto  
[e^{\si} e^{2\widetilde{\si}} + e^{\si} e^{2\widetilde{\si}}]e^{-\phi} e^{-\widetilde{\phi}} 
\Psi^m \widetilde{\Psi}_{m}  |0\rb_{\mathrm{NS}}\,.
\ee

\noindent
Using

\bea
&&e^{\si} e^{2\widetilde{\si}}e^{-\phi} e^{-\widetilde{\phi}} \Psi^m \widetilde{\Psi}_{m} 
= e^{\si} e^{-\phi} e^{2\widetilde{\si}} e^{-\widetilde{\phi}} \psi^a \widetilde{\psi}_{a}
+ \cdots \propto \la^+ (\widetilde{\la}^+)^2 \tha \ttha^4  +\cdots \,, \non \\
&&e^{2\si} e^{\widetilde{\si}}e^{-\phi} e^{-\widetilde{\phi}} \Psi^m \widetilde{\Psi}_{m} 
= e^{2\si} e^{-\phi} e^{\widetilde{\si}} e^{-\widetilde{\phi}} \psi_a \widetilde{\psi}^{a}
+ \cdots \propto (\la^+)^2 \widetilde{\la}^+ \tha^4 \ttha  +\cdots \,, \non 
\eea

\noindent
we find evidence for the remaining two terms in (\ref{bs0}).

Of course we do not pretend to have actually derived any of the terms 
in (\ref{bs0}) from the RNS expressions. Nevertheless, the above relations are 
an indication that terms of this form are expected to occur 
from a more careful treatment. 
It may be possible to make this more precise using the map 
in \cite{Berkovits:2001a}, although we have not 
tried to do so. We expect the map between the RNS and pure spinor 
boundary states to be rather subtle because of the different 
large/small Hilbert spaces involved in the two cases 
(\textit{cf.} the discussion in section~\ref{sMes}). Another likely source 
of complications is that the zero mode part of the RR boundary state 
is really more subtle than we pretended above, 
see, \textit{e.g.}, \cite{Yost:1988,Billo:1998a} for further discussions.

\subsection{The Generalisation to Lower Dimensional D$p$--Branes}
\label{lowerD}

As is well known the lower dimensional D--branes are obtained by 
imposing Dirichlet instead of Neumann conditions in some of 
the spacetime directions. 
In the general case, the boundary condition involving $x^m$ 
is (here $\partial_{\pm} = \partial_{\tau} \pm \partial_{\si}$)

\be
\left. \left( \partial_{-} x^m 
+ {R^{m}}_{n} \, \partial_{+} x^n \right)
\right|_{\tau=0} | B \rangle = 0\,,  \\
\ee

\noindent
where $R$ is the matrix

\begin{equation}
{R^{m}}_{n} = \matrto{\delta^a_{\ b}}{0}{0}{-\delta^i_{\ j}},
\end{equation}

\noindent
with $a, b = 0, \ldots, p$ and $i, j = p+1, \ldots, 9$.
Our ansatz for the boundary conditions in the spacetime fermionic sector is 
(for simplicity we only consider the type IIB theory; the type IIA 
case is similar)

\be  
\left. \left( p_\alpha \mp {R_{\alpha}}^{\beta} \, 
\widetilde{p}_\beta \right) \right|_{\tau=0} | B \rangle = 0, 
\quad
\left. \left( \theta^\alpha \pm {R^{\alpha}}_{\beta} \ttha^\beta  
\right) \right|_{\tau=0} | B \rangle = 0\,.
\ee

\noindent
At this point the matrices ${R_{\alpha}}^{\beta}$ and
${R^{\alpha}}_{\beta}$ should be seen as independent, but consistency 
of the above two equations gives the restriction 
${R^{\gamma}}_{\alpha} {R_{\gamma}}^{\beta} = \delta^{\beta}_{\alpha}$. 
Since this is just a relation between matrices, one also has
${R_{\alpha}}^{\gamma} {R^{\beta}}_{\gamma} = \delta^{\beta}_{\alpha}$. 

Further restrictions follow from the boundary conditions on the
supersymmetry and Lorentz currents:

\begin{equation}
\left( q_\alpha \mp {R_{\alpha}}^{\beta}
\widetilde{q}_{\beta} \right) \Big|_{\tau=0} \Bk = 0 \,,
\qquad
\left. \left( M^{m n} + {R^{m}}_{r} {R^{n}}_{s} \,
\widetilde{M}^{r s} \right) \right|_{\tau=0} | B \rangle = 0\,.
\end{equation}

\noindent
These conditions imply the relations 

\begin{equation}
\label{meq2}
\gamma^m_{\alpha \beta} {R^{\beta}}_{\gamma} {R_{m}}^{n}
=  {R_{\alpha}}^{\delta} \gamma^n_{\delta \gamma}\,,
\end{equation}

\noindent
and 

\begin{equation}
\label{meq3}
\gamma^m_{\alpha\beta} R^\beta{}_{[\si} 
{R^{\eta}}_{\rho]} \left( \gamma_m \right)_{\eta \delta} {R^{\delta}}_{\ka}  
=  {R_{\alpha}}^{\beta} \gamma^n_{\beta [\si} 
\left( \gamma_n \right)_{\rho] \ka}\,,
\end{equation}

\noindent
as well as

\begin{equation}
\label{meq4}
{R_{\alpha}}^{\eta} {\left( \gamma^{m n}
\right)^{\alpha}}_{\beta} {R^{\beta}}_{\delta} = 
 {R^{m}}_{r} {R^{n}}_{s} {\left( \gamma^{r s}
\right)^{\eta}}_{\delta}\,.
\end{equation}

\noindent
The solution to the above conditions is (for type IIB with $p$ odd) 

\begin{equation} \label{Rpsol}
{R_{\alpha}}^{\beta} = 
{\left( \gamma_{p+1 \ldots 9}
\right)_{\alpha}}^{\beta} = (-1)^{\left[\frac{p-1}{2}\right]} 
{\left( \gamma_{p+1 \ldots 9} \right)^{\alpha}}_{\beta}\,, \qquad 
{R^{\beta}}_{\al} = {\left( \gamma_{p+1 \ldots 9}
\right)^{\beta}}_{\al}  \,.
\end{equation}

\noindent
For $p=9$ both $R_{\alpha}{}^{\beta}$ and ${R^{\alpha}}_{\beta}$ are 
equal to $\de_\al^\beta$. One can check that the matrices 
(\ref{Rpsol}) satisfy all of the above restrictions. 
In the ghost sector the boundary conditions are 

\begin{equation}
\label{ghostconds}
\left. \left( \la^\alpha \pm {R^{\alpha}}_{\beta} \tla^\beta  
\right) \right|_{\tau=0} | B \rangle = 0\,, \qquad
\left( N^{m n} + {R^{m}}_{r} {R^{n}}_{s}
\widetilde{N}^{r s} \right) \Big|_{\tau=0} \Bk = 0\,,
\end{equation}

\noindent
which is consistent with $(Q+\widetilde{Q})|_{\tau=0} \Bk = 0$.

The boundary conditions for the modes of the fields can easily be worked out,
for instance one has (note the extra condition for the center of mass in 
the directions transverse to the brane)

\begin{eqnarray}
&
\left( \alpha^m_k + {R^{m}}_{n} 
\widetilde{\alpha}^n_{-k} \right) | B \rangle = 0\,,
\quad
X^i | B \rangle = y^i | B \rangle , \ i=p+1 ,..., 9\,,
 & \\
&
\left( p_{\alpha,k} \mp {R_{\alpha}}^{\beta} 
\widetilde{p}_{\beta,-k} \right)
 | B \rangle = 0\,, 
\quad
\left( \theta_k^\alpha \pm {R^{\alpha}}_{\beta} \tha^{\beta}{}_{-k} 
\right)  | B \rangle = 0\,, & \\
&
\left( L_k^{m n} + {R^{m}}_{\rho} {R^{n}}_{s}
\widetilde{L}_{-k}^{r s} \right)  | B \rangle = 0\,, &
\end{eqnarray}

\noindent
for any $k \in \mathbb{Z}$. The part of the prefactor in the boundary state involving 
the bosonic modes becomes  

\begin{equation}
 \delta^{9-p} ( X^i - y^i )
\exp \left( - \sum_{k=1}^\infty \frac{1}{k} 
\alpha_{-k}^m  R_{m n} \, \widetilde{\alpha}_{-k}^n \right) \,.
\end{equation}

\noindent
The prefactor involving the non--zero mode in the $p\theta$ sector is 

\begin{equation}
\exp \left( \mp \sum_{k=1}^\infty \left[ 
 p_{\alpha,-k} {R^{\alpha}}_{\beta} \, \ttha^{\beta}{}_{-k} 
-  \theta_{-k}^\alpha {R_{\alpha}}^{\beta}\, 
\widetilde{p}_{\beta,-k} \right] \right) ,
\end{equation}

\noindent
with a similar modification for the part involving the non--zero modes 
for the ghost fields. Finally, the zero mode part (\ref{bs0cov}) gets modified to 

\bea 
\Bk_{0} &=& \big[ U_{2,3}^{\al} R_\al{}^{\beta}\, \widetilde{U}_{1,2;\beta} 
\pm U^m_{2,4} R_{mn} \widetilde{U}^n_{1,1} 
+ U_{2,5;\al} R^{\al}{}_{\beta} \widetilde{U}^{\beta}_{1,0} \non \\
&& \;  \pm U_{1,2;\al} R^{\al}{}_\beta \widetilde{U}_{2,3}^{\beta} 
+  U^m_{1,1} R_{mn}  \widetilde{U}^n_{2,4} 
\pm U^\al_{1,0} R_{\al}{}^\beta \, \widetilde{U}_{2,5;\beta}
 \big] | Z \rb\,,
\eea

\noindent
where $| Z \rb$ is now also modified to reflect the new boundary conditions.

\setcounter{equation}{0}
\section{Applications and Checks} \label{secapp}

Although we have presented concise arguments in support of our expression for the boundary state in the pure spinor superstring, it is very important 
to check that results obtained from earlier calculations in the RNS superstring 
are recovered in this new formalism. One such test is to calculate scattering amplitudes in the presence of a D--brane and check that the earlier known results are correctly reproduced. As we shall see this requires some further work. Another test of our expression is to extend it to include the coupling to the gauge field living on the D--brane world--volume and to check that the resulting D--brane low--energy effective action is in agreement with the well--known results.

\subsection{Calculating Scattering Amplitudes} \label{scatt}

In the RNS approach, tree level scattering amplitudes in the presence of 
a D--brane can be calculated using (in this subsection we will only consider 
scattering amplitudes involving the bulk modes)

\be
\lb \cU | \left[ c_0  - \widetilde{c}_0 \right] \prod_{i=1}^{n-1} 
\int_{|z_i|>1} \!\!\!\!\! \D^2 z_i \, V_i(z_i,\bar{z_i})  \Bk \,.
\ee

\noindent
Here $\lb \cU | \equiv \lb 0|\cU(0,0)$, where the unintegrated 
vertex operator $\cU(z,\bar{z}) = U(z)\widetilde{U}(\bar{z})$ 
is assumed to have ghost number $(+1,+1)$ and to be in the appropriate 
picture to give a non--zero answer for the amplitude. The integrated 
vertex operators $V_i$ are assumed to have zero ghost-- and picture--numbers.
See, \textit{e.g.}, \cite{Craps:2000} for a discussion about how expressions 
such as the one above arise. The explicit $c$--ghost zero mode insertion 
is needed in order to get a non--zero answer. 
In the RNS case one does not have to be precise about the $c$--ghost 
insertions since the ghosts are Lorentz scalars and decouple completely. 
However, in the pure spinor formalism the $\la^\al$ ghosts have 
a Lorentz spinor index and do not decouple, so for later reference it 
is helpful to keep the ghosts in the above expression.

In generalising the above expression to the pure spinor case, one possible 
approach is to try to find an analogue for the $c_0$--ghost insertion. 
That additional zero mode insertions are needed also in the pure spinor case is clear from our expression for the boundary state (\ref{bs0cov}), along with  
the ghost charges and number of $\tha$'s in the unintegrated 
vertex operators, see, \textit{e.g.}, (\ref{UNS}), (\ref{UC}). 
One proposal for the required zero mode insertion 
is (as in RNS we will use the notation $c_0$; hopefully this will not lead to 
confusion) 

\be \label{c0}
c_0 = \half \oint (\la \ga^m\tha) (\tha \ga_{mnp}\tha) L^{np} 
\ee

\noindent
and analogously for $\widetilde{c}_0$. Here we have written an expression 
which, after performing the integral, involves all the modes 
of $\tha$, $\la$, $\ldots$, but in 
practice we only need the part involving purely the zero modes. 
Some evidence for the choice (\ref{c0}) can be obtained by noting that 
in RNS (using $\U(5)$ notation) 

\be
\ep_{abcde} \oint (\la^+ \tha^c \tha^d \tha^e (\psi^a\psi^b)) \propto 
\oint  (c\,(\psi^a\psi^b) (\psi_a \psi_b)) \propto \oint \D z \frac{c}{z^2} 
+ \cdots = c_0 + \cdots \,.
\ee

\noindent
Since one expects that $\psi^a \psi^b$ gets replaced by $L^{ab}$ 
in the pure spinor formalism and since one has 
$\ep_{abcde} \la^+ \tha^c \tha^d \tha^e 
\propto (\la \ga^m \tha)\, (\tha \ga_{m a b} \tha) + \cdots$, the 
covariant expression suggested in (\ref{c0}) seems like 
a reasonable guess.

With some work, it can be shown that the action of the $c_0$ operator 
on the zero mode states listed in table 2 is such that 

\be \label{corais}
c_0: U_{1,i} \rar - (-)^{\de_{i0}}U_{2,i+3}\,, \qquad i=0,1,2\,,
\ee

\noindent
\textit{i.e.}, it takes a vertex operator with ghost number 1 into a vertex operator with ghost number 2. Notice that in the present case what we call $c_0$ arises from integration of a conformal weight one field, unlike the $c_0$--ghost in the RNS case. The rather remarkable result (\ref{corais}) can be used to rewrite the zero mode part of the boundary state (\ref{bs0cov}) as (here we have also included an overall normalisation factor) 

\bea 
\Bk_{0} &=& \pm T_9 ( c_0 \mp  \widetilde{c_0}) \big[ -U_{1,1} \widetilde{U}_{1,1} 
\pm (U_{1,0} \widetilde{U}_{1,2} - \widetilde{U}_{1,0} U_{1,2} )
\big] \! |Z \rb \non \\
       &=&  T_9 \sum_{i=0}^{2} \big[ (\pm)^i U_{2,i{+}3} 
       \widetilde{U}_{1,2{-}i} + (\pm)^{i+1}
       U_{1,2{-}i} \widetilde{U}_{2,i{+}3} \big] |Z \rb \,.
\eea

\noindent
Here the factor $( c_0 \mp  \widetilde{c_0})$ is assumed to act only on the $U\widetilde{U}$ part (and not on $|Z\rb$).  
Thus we see that, apart from the $(c_0 \mp \widetilde{c}_0)$ factor, the 
zero mode boundary state is simply the sum of a ``graviton''--type and 
a ``$C$--field''--type vertex operator, \textit{cf.} (\ref{UNS}) and (\ref{UC}).  
This result is similar to the situation occurring in the RNS superstring, 
where an analogous result holds.

The above considerations lead to our proposal for the calculation 
of scattering amplitudes in the presence of a D--brane, 
in the pure spinor superstring:

\be \label{scpresc}
\lb \cU |( c_0 \pm  \widetilde{c_0})  \prod_{i=1}^{n-1} 
\int_{|z_i|>1} \!\!\!\!\! \D^2 z_i \, V_i(z_i,\bar{z_i})  \Bk \,,
\ee

\noindent
with 
\be 
\lb \cU |  = \lb Z | U \widetilde{U} \,,
\ee

\noindent
where $U \widetilde{U}$ is taken as a ghost number $(-1,-1)$ 
vertex operator and where by definition  
$\lb Z | = \vacb \prod_{I=1}^{11} Y^I(\tha\mp\ttha,\la\mp\tla)$. 
These $Y^I$ insertions are needed to give the right number 
of delta functions (when combined with the corresponding 
$Y^I$'s in $\Bk$). The choice of $\pm$ in (\ref{scpresc}) is required 
to get a non--vanishing amplitude. 
Instead of the above arguments, the prescription (\ref{scpresc}) 
can also be regarded as a definition.

Before turning to explicit calculations based on (\ref{scpresc}), 
let us make some comments. 
Since the operator we propose in (\ref{c0}) is the analogue 
of $c_0$ in the RNS superstring, 
one may naturally ask how it relates to the zero mode 
of the $b$--ghost, in its pure spinor version~\cite{Berkovits:2004a,Oda:2004}. 
There is no covariant expression for $b_0$ itself. However, 
in \cite{Berkovits:2004a} it was shown that there is a covariant expression 
for a ``picture--raised'' version of $b_0$. Moreover, in 
\cite{Anguelova:2004} it was further shown that even though $b_0$ itself is 
not covariant, the action of $b_0$ on $U = \la^\al \Phi_{\al}$ can be written covariantly as (modulo BRST exact terms)

\be
b_0: \la^\al \Phi_\al \rar G^\al \Phi_{\al} + H^{\al\beta} D_{\al} 
\Phi_{\beta} + K^{\al\beta\ga}D_{\ga}D_{\beta}\Phi_{\al} 
+ L^{\al\beta\ga\de} D_{\de}D_{\ga}D_{\beta}\Phi_{\al} \,.
\ee

\noindent
Concentrating on the zero mode vertex operator $U_{2,i+3}$, and 
using the schematic relations (here we only list the zero mode parts) $H^{\al\beta} \sim d d$, $K^{\al\beta\ga} \sim N d + J d$ and $L^{\al\beta\ga\de} \sim NN+JN+JJ$~\cite{Berkovits:2004a,Anguelova:2004},  
we see that the effect of $b_0$ is to remove 
three $\tha$'s and one $\la$ from $U_{2,i+3}$. Thus, it is reasonable to 
expect that $b_0$ maps $U_{2,i+3}$ to $U_{1,i}$ (modulo a constant). 
If this is indeed true then, at least when acting on these particular states, 
it is also possible that $b_0$ has the inverse action to $c_0$, as one would expect. 
Unfortunately, it is not possible to check this in detail 
as $H^{\al\beta}$, $K^{\al\beta\ga}$ and $L^{\al\beta\ga\de}$ have 
not yet been explicitly determined. 
 
Let us now consider a couple of examples, starting with the absorption 
of a single massless mode by a D$9$--brane. The vertex operator for the 
graviton is of the form $U^n_{1,1} h_{mn} \widetilde{U}^n_{1,1}$, \textit{cf.} (\ref{UNS}). The action of $c_0 \pm \widetilde{c}_0$ in (\ref{scpresc}) turns this vertex operator into $-\cU^h_{2;1} \mp  \cU^h_{1;2} 
\equiv - U^n_{2,4} \widetilde{U}^n_{1,1} h_{mn} 
\mp  U^n_{1,1} \widetilde{U}^n_{2,4}h_{mn}$ and 
the scattering amplitude becomes

\be \label{Vh1}
- \lb \cU^h_{2,1}\Bk_0 \mp \lb \cU^h_{1,2}\Bk_0 
\propto - T_9 \int \! \D^{10}x \, \tr(h)\,,
\ee

\noindent
which is zero, \textit{i.e.}, the 
correct result (in order to get a non--zero answer one needs to consider 
lower dimensional D--branes, in which case one finds $\tr(R \cdot h)$ instead of 
$\tr(h)$, which is also the known result). Notice that since the vertex operators we are dealing with involve zero modes only, the non--zero modes in the boundary state give no contribution. As another example, the vertex operator for a constant RR potential~\cite{Cornalba:2002} is proportional to
$U_{1,0} \, C\, \widetilde{U}_{1,2} 
+ \widetilde{U}_{1,0}\, \widetilde{C} \, U_{1,2}$, \textit{cf.} (\ref{Cvertex}).  
The action of $c_0 \pm \widetilde{c}_0$ in (\ref{scpresc}) turns this into 
$ \cU^C_{2;1} \pm \cU^C_{1;2} \equiv 
[    U_{2,3} \, C \, \widetilde{U}_{1,2} 
- \widetilde{U}_{1,0} \, \widetilde{C} \, U_{2,5} ]
\pm [ \widetilde{U}_{2,3} \, \widetilde{C} \, U_{1,2}
- U_{1,0} \, C \, \widetilde{U}_{2,5} ]$ and
the scattering amplitude becomes 

\be \label{VC1}
 \lb \cU^C_{2;1}\Bk_0 \pm \lb \cU^C_{1;2}\Bk_0 
\propto \pm T_9 \int \! \D^{10}x \, \tr(C - \widetilde{C}) 
\propto \pm T_9 \int \! C_{10}\,,
\ee

\noindent
which again is the correct result. Notice that the $\pm$, describing brane and anti--brane, also works out correctly: from (\ref{VC1}) follows that brane and anti--brane have opposite RR charges, whereas from (\ref{Vh1}) 
follows that they have the same tension.

One may also investigate the case with one additional vertex operator 
insertion. For the scattering involving two vertex operators for the graviton, 
or $B$--field, this is simple since the integrated vertex operators with 
zero ghost charge are very similar in both the RNS and pure spinor cases. 
Indeed, both can be written as 

\be
\ze_{mn} \int \! \D^2 z \, (\pa X^m + i k_r L^{rm} )
(\bar{\pa} X^n + i k_s L^{sn}) e^{ik\cdot X} \,,
\ee 

\noindent
the only difference being that $L^{mn} = -\Psi^m \Psi^n$ 
in the RNS case, whereas it is $L^{mn} = M^{mn} + N^{mn}$ in the pure spinor case.
In the pure spinor case there are also subleading terms 
in the vertex operator, at higher order in the $\tha$ expansion, but these 
will not contribute to the scattering amplitude as they give rise 
to expressions with too many $\tha$ zero modes. Since $L^{mn}$ 
and $-\Psi^m\Psi^n$ have the same boundary conditions and the same OPE's 
with the unintegrated vertex operators for the massless modes 
(this follows immediately from the Lorentz index structure), 
the scattering amplitudes will necessarily be the same. Thus, we find agreement 
with the corresponding RNS results, obtained, 
\textit{e.g.}, in~\cite{Gubser:1996,Garousi:1996}. Consequently, we 
also find agreement with the terms in the D--brane effective action 
extracted from these results, see, \textit{e.g.}, \cite{Bachas:1999}.

Above we mostly discussed the scattering of massless modes, but it should 
also be possible to consider the massive modes. Scattering amplitudes 
in the presence of a D--brane beyond tree level might also be within 
reach, using the results in \cite{Berkovits:2004a}.

\subsection{Coupling to a Gauge Field} \label{Ffield}

In this subsection we incorporate the gauge field living on the world--volume 
of the D--brane into the previous analysis. For simplicity we consider the case 
of a D$9$--brane with a constant gauge field, and will comment on the more 
general case later. The boundary condition involving $x^m$ is of course the same as 
in the RNS superstring:  

\be
\left. (\partial_{\tau} x^m - f^{m}{}_n \, \partial_{\si}x^n ) 
\right|_{\tau=0} | B(f) \rangle = 0\,.  
\ee

\noindent
This can also be written as (here $\pa_{\pm} = \pa_{\tau} \pm \pa_{\si}$) 

\be
\left. (\pa_{-} x^m + R(f)^{m}{}_{n}\, \pa_{+} x^n) \right|_{\tau=0} | B(f) \rangle=0\,, 
\quad 
\mathrm{where} \quad R(f)^{m}{}_{n}= \bigg(\frac{1-f}{1+f}\bigg)\!\!{{\phantom{\big|}}^{m}}_{n}\,.
\ee

\noindent
Since the Lorentz current and the supersymmetry generator are 
physical quantities, it must be the case that their boundary conditions should 
agree with the ones in the RNS superstring~\cite{Callan:1988}, \textit{i.e.}, one must require

\be  \label{bcFLq}
(L^{mn} +  R(f)^{m}{}_{r} R(f)^{n}{}_{s}\widetilde{L}^{rs}) | B(f) \rangle=0 \,, \qquad 
 (q_{\al} \mp R(f)_{\al}{}^{\beta} \widetilde{q}_{\beta})| B(f) \rangle=0\,.
\ee

\noindent
Here $R(f)_\al{}^\beta$ should agree with the RNS result~\cite{Callan:1988}

\be
R(f)_\al{}^\beta = \det(1+f)^{-1/2} \
\mbox{\AE}(- \ga_{mn}f^{mn})_\al{}^\beta\,,
\ee

\noindent
where $\mbox{\AE}$ is the antisymmetrised exponential

\be
\mbox{\AE}(- \ga_{mn}f^{mn}) = \sum_{p=0}^5 {\ts \frac{1}{p!} }(-)^p
\ga_{m_1 n_1 \cdots m_p n_p} f^{m_1 n_1}\cdots f^{m_p n_p}\,.
\ee

\noindent
Note that $R(f)_{\al}{}^{\beta} = R(-f)^{\beta}{}_{\al} $. 
Furthermore we have the relations

\be \label{Rrels}
R(-f) = R(f)^{-1} \,, \qquad R(f)^{-1} \ga^m R(f) 
=\left[\frac{1-f}{1+f}\right]\!\!{{\phantom{\big|}}^{m}}_{n} \ga^n\,.
\ee

\noindent
Now, the following boundary conditions  

\bea
&& (\tha^\al \pm \ttha^\beta R(-f)_{\beta}{}^{\al})  | B(f) \rangle =0\,, 
 \qquad (p_\al \mp R(f)_{\al}{}^{\beta}\tp_\beta) | B(f) \rangle=0 \,, \non \\
&& (\la^\al \pm \tla^\beta R(-f)_{\beta}{}^{\al}) | B(f) \rangle=0\,,
\eea

\noindent
are consistent with (\ref{bcFLq}) and reduce to the correct 
boundary conditions when $f_{mn}=0$. Furthermore, they imply that the 
boundary conditions involving $Q$, $g$ and $h$ are independent 
of $f_{mn}$ as they should be, and also that, \textit{e.g.}, the boundary condition 
involving $\la \ga^m \, \tha$ has the same structure as the one involving 
$\pa_{-} x^m$. Note also that the above boundary condition involving 
$\la^\al$ and $\tla^\al$ is consistent with the pure spinor conditions because of 
the relations (\ref{Rrels}). An alternative 
approach one could follow, instead of using 
the known boundary conditions from the RNS superstring, is to proceed as in section \ref{lowerD} and make an ansatz for the boundary conditions involving $\tha^\al$, $p_\al$ and $\la^\al$. 
Consistency of the other boundary conditions then leads to the same 
result as above for $R(f)_\al{}^\beta$.

The prefactor part of the boundary state involving the non--zero modes 
is closely analogous to the one in sections~\ref{bnzm}, \ref{lowerD}. 
The part involving the modes of $\partial x^m$ is  

\begin{equation}
\exp \left( - \sum_{k=1}^\infty \frac{1}{k} 
\alpha_{-k}^m  R_{m n}(f) \, \widetilde{\alpha}_{-k}^n \right) ,
\end{equation}

\noindent
whereas the prefactor involving the non--zero modes in the $p\theta$ sector is 

\begin{equation}
\exp \left( \mp \sum_{k=1}^\infty \left[ 
 p_{\alpha,-k} {R^{\alpha}}_{\beta}(f) \, \ttha^{\beta}{}_{-k} 
-  \theta_{-k}^\alpha {R_{\alpha}}^{\beta} (f)\, 
\widetilde{p}_{\beta,-k} \right] \right) .
\end{equation}

\noindent
A convenient basis for calculations is to choose $f$ to be block--diagonal 
with the $2{\times}2$ blocks along the diagonal proportional to 
{\footnotesize $\left( \!\! \ba{cc} 0& f_i \\ -f_i & 0 \ea \!\! \right)$} 
with $i=1,\ldots,5$. In this basis ${R^{\alpha}}_{\beta}(f)$ takes the form 

\be 
\frac{(1 - i f_1 \si_3 ) }{\sqrt{1+ f_1^2}} \otimes \cdots 
\otimes \frac{(1 - i f_5 \si_3 )}{\sqrt{1+f_5^2}}\,.
\ee

\noindent
Thus the boundary conditions can be written in the $\U(5)$ notation as 
(see appendix \ref{u5} for more details on the $\U(5)$ notation for spinors)

\bea
\bigg(\tha^{\pm\pm\pm\pm\pm} \pm 
\frac{(1 \mp i f_1)(1\mp i f_2)(1\mp i f_3)(1\mp i f_4)(1 \mp i f_5)}{
\sqrt{1+f_1^2} \sqrt{1+f_2^2}\sqrt{1+f_3^2}\sqrt{1+f_4^2}\sqrt{1+f_5^2}}
\,\, \ttha^{\pm \pm \pm \pm \pm}\bigg) | B(f) \rangle\,, \non \\
\bigg(p_{\pm\pm\pm\pm\pm} \mp
\frac{(1 \pm i f_1)(1\pm i f_2)(1\pm i f_3)(1\pm i f_4)(1 \pm i f_5)}{
\sqrt{1+f_1^2} \sqrt{1+f_2^2}\sqrt{1+f_3^2}\sqrt{1+f_4^2}\sqrt{1+f_5^2}}
\,\, \tp_{\pm \pm \pm \pm \pm}\bigg)| B(f) \rangle\,,
\eea

\noindent
with exactly the same form for the boundary condition involving $\la$, $\tla$ 
as for the one involving $\tha$, $\ttha$. 
In this form we again note that the boundary condition is consistent with the 
pure spinor constraints, as can be seen from the explicit solution of 
the constraint in the $\U(5)$ basis, \textit{cf.} (\ref{lapar}).
In the $\U(5)$ basis the part of the prefactor in the boundary state 
involving the non--zero modes for the ghost fields is particularly simple 
since it can, \textit{e.g.}, be written using only  the conjugate 
pairs $(w_+,\la^+)$, $(w^{ab},\la_{ab})$ 
and the right--moving partners.

The zero mode part of the boundary state which satisfies 
the above boundary conditions is 

\bea \label{B0F}
\!\! |B(f) \rangle_{0} \!\!\!&=&\!\!\! T_9 \sqrt{\det(1+f)}
\big[ U_{2,3}^{\al} R(-f)_\al{}^{\beta} \, \widetilde{U}_{1,2;\beta} 
\pm U^m_{2,4} R(f)_{mn} \widetilde{U}^n_{1,1} 
+ U_{2,5;\al} R(f)^{\al}{}_{\beta} \widetilde{U}^{\beta}_{1,0} \non \\
&& \;  \pm U_{1,2\al} R(f)^{\al}{}_\beta \widetilde{U}_{2,3}^{\beta} 
+  U^m_{1,1} R(f)_{mn}  \widetilde{U}^n_{2,4} 
\pm U^\al_{1,0} R(-f)_{\al}{}^\beta \, \widetilde{U}_{2,5;\beta}
 \big] |Z \rb\,.
\eea

\noindent
As in earlier sections, we have suppressed the $Y^I$'s hidden in $|Z \rb$. 
In the present case the $Y^I$'s seemingly have an $f_{mn}$--dependence. 
However, this $f$--dependence can be 
absorbed into the $C_\al^I$ and $\widetilde{C}_{\al}^I$'s and therefore 
does not give rise to any additional $f$--dependence when calculating 
correlation functions or scattering amplitudes. In (\ref{B0F}) we have included 
an $f$--dependent overall normalisation factor needed to find agreement with 
the usual D--brane effective action (see below). It may be possible to 
understand this normalisation factor from a more careful analysis.

Having determined the dependence on the 
world--volume gauge field in the boundary state, we can now use the resulting 
expressions to calculate scattering amplitudes and, for instance, derive 
the D--brane low--energy effective action. However, before turning to this 
problem, let us discuss some puzzling aspects not touched upon above.
In fact, we have obtained the boundary conditions using the closed string language, 
but one should also be able to derive these boundary conditions directly in 
the pure spinor superstring, starting from the pure spinor 
superstring sigma model (\ref{flat-action}) plus a boundary action. 
Such an analysis was carried out in \cite{Berkovits:2002c}. 
In that paper, the resulting boundary conditions 
were expressed in superfield notation. 
Consistency requires that the superfields satisfy certain constraint 
equations. These equations put the superfields on--shell and were also obtained 
in \cite{Kerstan:2002} using the 
superembedding approach for D--branes. 
Moreover, it was shown in \cite{Kerstan:2002,Berkovits:2002c} that 
these equations have 16 (manifest) linear and also 16 (non--manifest) 
non--linear supersymmetries and that the superfield equations therefore 
have to imply the (super) DBI equations of motion. 

To make contact with the boundary conditions above (for a constant 
gauge field strength) one has to expand the superfield equations in 
components. It may seem strange that in order to obtain the boundary conditions 
involving just the gauge field one first 
needs to ``solve'' the superfield equations (\textit{i.e.}, determine the component 
expansion), since these equations also encode the equations of 
motion for the DBI action. It thus might seem that in order to determine the boundary 
state (which can then be used to obtain the DBI action) one first has 
to obtain the equations of motion of the DBI action. However, this is not quite so. Rather, in order to find 
the boundary conditions for a constant gauge field strength one actually 
does not need 
to solve the superfield equations completely, but only determine the first few  
components in the $\tha$ expansion and the (algebraic) relations 
between them. This will be further discussed in what follows.

Based on comparison with results in the RNS superstring 
(see, \textit{e.g.}, \cite{Hashimoto:1999}) one expects that 
it should also be possible to write the boundary state in the form  
$e^{S_\mathrm{b}(f)} |B(f=0) \rb$, where $S_\mathrm{b}(f)$ is essentially 
the boundary action. Naively, one would expect 
$S_\mathrm{b}(f) \propto \int \D\si (\pa_\si X^m a_m(X) 
+ \half L^{mn} f_{mn})$ since this is the form of 
$S_\mathrm{b}(f)$ in the RNS boundary state (see, \textit{e.g.}, \cite{Hashimoto:1999}). Here $L_{mn}$ is the boundary value of 
the RNS Lorentz current and one might think that the only modification one 
needs to do is to replace the RNS Lorentz current by its pure spinor analogue. 
However, this naive ansatz does not work (as was observed 
in~\cite{Tseytlin:1998} in 
the related context of light--cone GS boundary 
states~\cite{Green:1996}).
In order to understand how to solve the problem, and also how to make contact with the result in~\cite{Berkovits:2002c}, let us focus on 
the $p\tha$ sector. The naive guess would be that $S_\mathrm{b}(f)$ involves 
$-\half \int \D\si f_{mn} (p \pm \widetilde{p}) \ga^{mn} (\tha \mp \ttha)$, in 
which case $S_\mathrm{b}(f)$ would simply be structurally the same as the 
vertex operator (\ref{fint}).  However, let us instead 
(as in~\cite{Tseytlin:1998}) 
consider the more general form where $S_\mathrm{b}(f)$ is 
$-\half \int \D\si  (p_\al \pm \widetilde{p}_\al) (\tha^\beta \mp \ttha^\beta) 
G_\al{}^\beta(f)$. 
This leads to the boundary condition 

\be
[(\tha^\beta \pm \ttha^\beta) + 
(\tha^\beta \mp \ttha^\beta) G_\al{}^\beta(f)] | B(f) \rangle \,.
\ee

\noindent
By comparison with the previous boundary condition we find agreement provided  
$G_\al{}^\beta$ is related to $R_{\al}{}^\beta$ via 
$G= (1-R)/(1+R)$. Using this relation 
we can determine $G_{\al}{}^{\beta}(f)$ using the known form of 
$R_\al{}^\beta(f)$. 
The explicit form is rather complicated but it is easy to show (\textit{e.g.}, using the block--diagonal basis for $f_{mn}$ to simplify the calculation) that 
for D$p$--branes with $p > 4$, $G_{\al}{}^{\beta}(f)$ contains more than 
just a $\ga_{(2)}$ part; in particular it also has a $\ga_{(6)}$ piece, 
meaning that $S_\mathrm{b}(f)$ does not just involve the ``Lorentz generators'' 
$(p \pm \widetilde{p}) \ga^{mn} (\tha \mp \ttha)$ as one naively would have 
expected. Furthermore, $G$ is of course in general not linear in $f_{mn}$. 

On the other hand, on general grounds one expects $S_\mathrm{b}(f)$ to 
be simply given by the boundary action constructed in \cite{Berkovits:2002c}. 
However, this raises a puzzle since in 
this action only the $\ga_{(2)}$ piece occurs. To see this we first note that 
the boundary action given in \cite{Berkovits:2002c} is written in terms of the fields $W^{\al}$ and $F_{mn}$ (which were defined in \cite{Berkovits:2002c} and are different from the Yang--Mills ones, given in appendix A). Keeping only the leading order terms in the $\tha$ expansion of $W^{\al}$ and $F_{mn}$, which depend non--linearly on $f_{mn}$, and using $W^{\al} \propto (\tha \mp \ttha)^\beta {G_\beta}^{\al}(f)$ 
together with  the relation $D_{\al} W^{\beta} = \frac{1}{4} F_{mn}(f) (\ga^{mn})_{\al}{}^\beta$ (which holds for the leading order terms we consider),  
one finds that the boundary action given in \cite{Berkovits:2002c} contains 
a term as above, but where $G_{\al}{}^{\beta}(f)$ only has a $\ga_{(2)}$ piece.
In fact, it is even easier to see that only a $\ga_{(2)}$ piece occurs in 
the ghost sector, either from the boundary action in \cite{Berkovits:2002c} 
or more directly from the boundary condition 
\cite{Berkovits:2002c} $\la^\al \pm \tla^\al 
= -\frac{1}{4}(\la^\beta \mp \tla^\beta) (\ga_{mn})_\beta{}^\al F_{mn}(f)$  
(there is a similar problem with the $\tha$, $\ttha$ boundary condition). 
Thus there seems to be a contradiction. However, this is easily 
resolved by replacing $(\ga_{mn})_\al{}^\beta F_{mn}(f)$ by a multiple of $G_{\al}{}^{\beta} (f)$ (not affecting any of the conclusions in \cite{Berkovits:2002c}). In fact, such a replacement is also required to get agreement with the results in \cite{Kerstan:2002} where the analogue of 
$(\ga_{mn})_\beta{}^\al F_{mn}(f)$ was denoted $h_\al{}^\beta(f)$ 
and was explicitly shown to contain more than just a $\ga_{(2)}$ 
piece (see also \cite{Akulov:1998}). Thus, after this replacement 
everything agrees. There is still a slight worry in the ghost sector since, as noted above, the boundary action seems to involve more than just 
the ``Lorentz generators'' (\textit{i.e.}, more than just $\ga_{(2)}$). This fact seems to be in conflict with the ghost gauge invariance $w_\alpha \to w_\alpha + ( \gamma^m \lambda)_\alpha \Lambda_m$, as the term coupling ghosts and gauge field is not invariant. However\footnote{We thank N. Berkovits for discussions on this issue.}, the complete action in \cite{Berkovits:2002c} is invariant under the ghost gauge transformation provided the above boundary conditions hold. This again amounts to replacing ${( \gamma^{mn} )_\alpha}^\beta F_{mn} (f)$ in \cite{Berkovits:2002c} with a multiple of our ${G_\alpha}^\beta (f)$, which is required for consistency with the pure spinor constraint (\textit{cf.}, the discussion at the beginning of this section).

We will now test our proposal (\ref{B0F}).
As in RNS \cite{DiVecchia:1997,Billo:1998a,DiVecchia:1999a}, the boundary 
state can be used to calculate terms in the D--brane effective action. 
In particular, using (\ref{B0F}) we will determine the $f$--dependence 
of the D--brane low--energy effective action. 
One difference compared to the RNS case is that in the pure spinor calculation 
only zero modes are involved. This makes the determination of the DBI 
and WZ parts of the actions very similar. 
More specifically, using the same method as in \cite{DiVecchia:1997,Billo:1998a,DiVecchia:1999a} we shall next determine the linear coupling of the D--brane  
to both the graviton and the RR $C$--field potential, and from the obtained result extract the D--brane effective action. 

The linear coupling (in the action) to the graviton $h_{mn}$ 
($g_{mn}=\de_{mn}+h_{mn}$) is obtained 
from (\textit{cf.} (\ref{Vh1}); see the discussion 
in the previous subsection)

\be
- \lb \cU^h_{1,2} |B(f) \rangle_{0} \mp  \lb 
\cU^h_{2,1} |B(f) \rangle_{0} 
\propto - T_9\int \!\! \D^{10}x \, \tr(h R(f) ) \sqrt{\det(1+f)}
\ee

\noindent
from which we deduce, following~\cite{DiVecchia:1997}, 
the DBI part of the D--brane effective action, 
$-T_9 \int  \! \D^{10}x \sqrt{\det(g+f)}$. 
The linear coupling (in the action) to the RR form fields 
$C_p$ is obtained from (\textit{cf.} (\ref{VC1}))

\bea
\lb \cU^C_{1,2} |B(f) \rangle_{0} \pm \lb \cU^C_{2,1} |B(f) \rangle_{0}  
&\propto& \pm T_9 \sqrt{\det(1+f)} \int \!\! \D^{10}x \, 
\tr(C R(f) - R(f)\widetilde{C} ) \non \\
&\propto& \pm T_9 \int \!\! C \wedge e^f \,,
\eea

\noindent
which we recognise as the WZ part of the action, 
$\pm T_9 \int C \wedge e^F$. 
There are several possible extensions of the above analysis  
which we shall leave for future work \cite{Schiappa:2005}.

\section{Final Remarks} \label{sdisc}

In this paper we have studied the pure spinor D--brane boundary state 
in the simplest setting: a flat spacetime background. 
Clearly, much work remains to be done and many extensions could be studied in the future. Ideally, one would like to address problems which are inaccessible 
using standard RNS methods. 

As we have discussed at length, the hardest issue in the understanding of the boundary state is its zero mode sector. Our main result for this part of the boundary state is given in (\ref{bs0}), or in slightly more condensed form in  
(\ref{bs0cov}). Our evidence for this result is that it was constructed in order to 
satisfy the correct boundary conditions. It further leads to the correct couplings to the bulk supergravity fields (\textit{i.e.}, one--point functions), and also correctly reproduces certain scattering amplitudes involving two vertex operator 
insertions. These tests did involve some further assumptions beyond just 
the boundary conditions; in particular in section \ref{scatt} we proposed 
a prescription for calculating scattering amplitudes in the presence of 
a D--brane. We have also coupled the boundary state to a gauge field 
(see section~\ref{Ffield}). The boundary conditions for the gauge field 
are consistent with the pure spinor constraints and the calculation 
of scattering amplitudes using the framework discussed in section~\ref{scatt} 
leads to expressions which agree with the known low--energy D--brane effective 
action for the world--volume gauge field. 

One would further like to check that 
$\Bb P |B\rangle$ (where $P$ is the propagator) is consistent with 
the known RNS result. 
This analysis is hampered by the difficulty of constructing  
the propagator in the pure spinor superstring. 
We note that in the RNS superstring the total amplitude vanishes after summing over the various NS and R sectors. 
However, a ghost zero mode insertion is needed to get a non--vanishing result 
within each sector.
In the pure spinor case it seems that the amplitude also vanishes, 
as required for consistency, but not as a result 
of cancellations between expressions involving all modes, rather as 
a consequence of the cancellation amongst the zero mode pieces alone.

\section*{Acknowledgements}
We would like to thank Troels Harmark and Marcos Mari\~no for collaboration 
in 2002 on some of the topics discussed in this paper. We would also like to thank Nathan Berkovits and Pietro Antonio Grassi for discussions and comments. RS would like to thank CECS (Valdivia) for very nice hospitality during the course of this work. RS is supported in part by funds provided by the Funda\c c\~ao para a Ci\^encia e a Tecnologia, under the grants SFRH/BPD/7190/2001 and FEDER--POCTI/FNU/38004/2001.

\appendix

\setcounter{equation}{0}
\section{Flat ${\mathcal{N}}=1$, $d=10$ Superspace} \label{appFlat}

The ${\mathcal{N}}=1$, $d=10$ superspace coordinates are $\left( x^{m},
\theta^{\alpha} \right)$. The supersymmetry transformations acting on 
superfields are generated by

\begin{equation}
\mathcal{Q}_{\alpha} = \frac{\partial}{\partial \theta^{\alpha}} -
\frac{1}{2}  \left( \gamma^{m} \theta \right)_{\alpha}
\frac{\partial}{\partial x^{m}}\,,
\end{equation}

\noindent
satisfying

\begin{equation}
\left\{ \mathcal{Q}_{\alpha}, \mathcal{Q}_{\beta} \right\} = -
\gamma^{m}_{\alpha\beta} \frac{\partial}{\partial x^{m}}\,.
\end{equation}

\noindent
The vector field $\frac{\partial}{\partial x^{m}}$ is invariant under the
supersymmetry transformations, as is the usual supersymmetric derivative,

\begin{equation}
D_{\alpha} = \frac{\partial}{\partial \theta^{\alpha}} + \frac{1}{2} \left( 
\gamma^{m} \theta \right)_{\alpha} \frac{\partial}{\partial x^{m}}\,,
\end{equation}

\noindent
which satisfies,

\begin{equation}
\left\{ D_{\alpha}, D_{\beta} \right\} = \gamma^{m}_{\alpha\beta}
\frac{\partial}{\partial x^{m}}\,.
\end{equation}

To describe ${\mathcal{N}}=1$ super Yang--Mills in superspace one introduces 
the superfield potentials $A_{\al}(x,\tha)$, $A_{m}(x,\tha)$, and their 
associated field strengths. Here $A_{m} (x,\theta) = a_{m}(x) + \cdots$,
where the field $a_{m} (x)$ is identified with the super Yang--Mills
gluon. As this construction is fairly well--known, we shall be brief.

An analysis of the superspace Bianchi identities leads to the following 
relations between the potentials and their field strengths

\bea \label{ssrels}
&&D_{\alpha} A_{\beta} + D_{\beta} A_{\alpha} =
 \gamma^{m}_{\alpha\beta} A_{m} \,, \non \\
&&  D_{\al} A_{m} - \partial_{m}
A_{\al}  = \left( \gamma_{m}
\right)_{\alpha\beta} W^{\beta} \,,  \\
&&\pa_m A_n - \pa_n A_m \equiv F_{mn}  = - {\ts \frac{1}{8} } 
{\left( \gamma_{mn}\right)^{\alpha}}_{\beta} D_{\alpha} W^{\beta}  \non \,.
\eea

\noindent
Multiplying the first equation with the gamma matrix $\ga_n^{\al\beta}$ 
yields the relation

\begin{equation} \label{AavsAal}
A_{n} (x,\theta) = {\ts \frac{1}{8} } \gamma_{n}^{\alpha\beta} D_{\alpha}
A_{\beta} (x,\theta)\,,
\end{equation}

\noindent
which defines $A_n$ in terms of $A_\al$. Note that the gauge 
invariance $A_{\alpha} \to A_{\alpha} 
+ D_{\alpha} \Omega$  implies the standard gauge invariance 
$A_{m} \to A_{m} + \partial_{m} \Omega$. Multiplying instead 
with $\gamma^{\al\beta}_{m_{1} \cdots m_{5}}$ yields 

\begin{equation}
\gamma^{\alpha\beta}_{m_{1} \cdots m_{5}} D_{\alpha} A_{\beta} = 0\,, 
\end{equation}

\noindent
which can be shown to imply the super Yang--Mills equations of motion, 
\textit{i.e.}, this superspace constraint puts the theory on--shell. 
Multiplying the second equation in (\ref{ssrels}) 
with $(\ga^m)^{\eta\al}$ yields

\be
W^{\eta} = {\ts \frac{1}{10}} (\ga^m)^{\eta\al} 
( D_{\al} A_{m} - \partial_{m} A_{\al})   \,,
\ee

\noindent
which together with (\ref{AavsAal}) defines $W^{\eta}$ in terms of $A_\al$. 
Here $W^{\eta} (x,\theta) = \xi^{\eta}(x) + \cdots$, 
where $\xi^{\eta}(x)$ is identified with the super Yang--Mills gluino field. 

For a super Yang--Mills background containing a gluon field $a_{m} (x)$ 
with constant field strength $f_{mn}$, and a constant gluino field 
$\xi^{\alpha}$, one can show that the $\tha$ expansions of the above 
superfields are

\begin{eqnarray}
 A_{\alpha} (x,\theta) &=& {\ts \frac{1}{2} } \left( \gamma^{m} \theta
\right)_{\alpha} a_{m} (x) +{\ts \frac{1}{3}} \left( \gamma^{m} \theta
\right)_{\alpha} \left( \theta \gamma_{m} \xi \right) - {\ts \frac{1}{32} }  
\left( \theta \gamma^{n p q} \theta \right) \left( \gamma_{n} \theta
\right)_{\alpha}
f_{p q}\,, \quad \\
A_{m} (x,\theta) &=& a_{m} (x) + \theta \gamma_{m} \xi 
- {\ts \frac{1}{8} } \left( \theta \gamma_{m n p} \theta \right) f^{n p}\,, \\
W^{\alpha} (x,\theta) &=& \xi^\al + {\ts \frac{1}{4} } 
\left(\theta \gamma^{mn} \right)^{\alpha} f_{mn}\,, \\
F_{mn} (x,\theta) &=& f_{mn}\,.
\end{eqnarray}

\setcounter{equation}{0}
\section{The $\mathrm{U}(5)$ Formalism} \label{u5}

It will be occasionally useful to temporarily break $\SO(10)$ to 
$\U(5) \approx \U(1){\times}\SU(5)$. Under this breaking pattern, the vector 
representation of $\SO(10)$ decomposes as $10 \rar 5 \oplus \bar{5}$. 
The components of a $\SO(10)$ vector $V^m$ are related to the 
components of the two $\U(5)$ 
representations $v^a$, $v_a$, according to 
$v^a = \frac{1}{2}( V^a + i V^{a+5})$ for 
the $5$ and $v_a = \frac{1}{2}(V^a - i V^{a+5})$ for the $\bar{5}$; here 
$a=1,\ldots,5$. Analogous expressions can be derived for a tensor with an
 arbitrary number of vector indices. 
The following representation for the $\U(5)$ components $(\ga^a)_{\al \bet}$ 
and $(\ga_a)_{\al\bet}$ of the $\SO(10)$ gamma matrices $\Ga^{m}_{\al\bet}$ 
is useful (see, \textit{e.g.}, \cite{Atick:1986})

\begin{equation}
\begin{array}{lcl}
(\ga^1)_{\al\beta} = -\frac{1+\si_3}{2} \otimes \si_2 \otimes \si_1 
\otimes \si_2 \otimes \si_1 \,, & \quad &  (\ga_1)_{\al\beta} = 
-\frac{1-\si_3}{2}\otimes \si_2 \otimes \si_1 \otimes \si_2 \otimes \si_1 \,, 
\\
(\ga^2)_{\al\beta} = -\si_2\otimes \frac{1+\si_3}{2}\otimes  \si_1 
\otimes \si_2 \otimes \si_1\,, &\quad &  (\ga_2)_{\al\beta} = 
-\si_2\otimes \frac{1-\si_3}{2} \otimes \si_1 \otimes \si_2 \otimes \si_1 \,, 
\\
(\ga^3)_{\al\beta} = -\si_2\otimes \si_1 \otimes \frac{1+\si_3}{2}
\otimes  \si_2 \otimes \si_1\,, & \quad&  (\ga_3)_{\al\beta} = 
-\si_2\otimes \si_1\otimes \frac{1-\si_3}{2}  \otimes \si_2 \otimes \si_1 \,,
\\
(\ga^4)_{\al\beta} = -\si_2\otimes \si_1 \otimes \si_2 \otimes 
\frac{1+\si_3}{2} \otimes \si_1\,, & \quad&  (\ga_4)_{\al\beta} 
= -\si_2\otimes \si_1\otimes\si_2\otimes \frac{1-\si_3}{2} \otimes \si_1 \,,
\\ 
(\ga^5)_{\al\beta} = -\si_2\otimes \si_1 \otimes \si_2 \otimes 
\si_1 \otimes \frac{1+\si_3}{2} \,, & \quad&  (\ga_5)_{\al\beta} 
= -\si_2\otimes \si_1\otimes\si_2\otimes \si_1 \otimes \frac{1-\si_3}{2} \,.
\end{array}
\end{equation}

\noindent
Indices are raised and lowered with $\ep^{\al\bet} = -\si_1\otimes 
\si_2\otimes \si_1\otimes \si_2 \otimes \si_1$ and its inverse 
$\ep_{\al\bet}$, according to the rule $T^{\al\bet} 
= \ep^{\al \de}T_{\de \rho} \ep^{\rho \bet}$. 
The above matrices satisfy $\{\ga^a,\ga_b\}=\de^a_b$. 
From this result it follows that the corresponding $\Ga^m$'s 
satisfy $\{\Ga^m,\Ga^n\} = 2 \eta^{mn}$. The above matrices are 
$32{\times} 32$ dimensional, but we will only consider the 
restricted action on 16 dimensional Weyl spinors. The restriction 
of $\Ga^m_{\al\bet}$ to this subspace will be denoted by $\ga^m_{\al\bet}$.  

A spinor of $\SO(10)$ is conveniently represented as the direct product of 
5 $\SO(2)$ spinors. Denoting the $\SO(2)$ spinor {\tiny $ 
\left( \!\!\! \ba{c} 1 \\ 0\ea \!\!\! \right)$ } by $+$ and  
{\tiny $\left( \!\!\! \ba{c} 0 \\ 1\ea \!\!\! \right)$ } by $-$, 
$\SO(10)$ spinors are naturally labelled by a composite index 
$(\pm,\pm,\pm,\pm,\pm)$, where all 32 possible choices are allowed. 
The above $\ga$ matrices act on this basis in the natural way. 
In this basis a 32 dimensional spinor splits into $16\oplus\overline{16}$. 
Spinors with an odd (even) number of $+$'s are Weyl (anti--Weyl) spinors. 
A sixteen dimensional spinor $\la^\al$ further decomposes as $16 \rar 
1 \oplus 5 \oplus \overline{10}$ (and similarly for the $\overline{16}$ 
representation). 
The components with one $+$ will be denoted $\la^a$ and belong to the $5$ 
representation; the components with three $+$'s will denoted by $\la_{ab}
=-\la_{ba}$ and belong to the $\overline{10}$ representation. Finally, the 
component with five $+$'s is the singlet and is denoted $\la^+$ (the 
difference between the number of $+$'s and $-$'s divided by 2 is the 
$\U(1)$ quantum number).

Using the above results one can derive the relations (our conventions 
differ slightly from those of Berkovits)

\begin{equation}
\la^\al (\ga^a)_{\al\bet} \la^\bet = 2[\la^+ \la^a - {\ts \frac{1}{8} } 
\ep^{abcde}\la_{bc}\la_{de}] \,, \qquad \la^\al (\ga_a)_{\al\bet} \la^\bet 
= 2 \la_{ab}\la^b \,,
\end{equation}

\noindent
which gives the parameterisation (\ref{lapar}). Define $w_\al$ via

\begin{equation} \label{w}
w_+ = e^{-s}(\pa t + a \pa s)\,, \qquad w^{ab} = -2 v^{ab} \,, \qquad w_a =0\,,
\end{equation}

\noindent
where $a$ is a normal ordering constant. 
One can readily check that $w_\al$ satisfies the following OPE with $\la^\bet$

\begin{equation} \label{wlOPE}
w_\al(y) \la^\bet (z) \sim \frac{\de_\al^\bet}{y-z} - \half  
\frac{(\ga_{m})^{\bet +}}{y-z}e^{-s}(\ga^{m})_{\al\de}\la^\de\,.
\end{equation}

\noindent
Note that the OPE's between $w_+$, $\la^+$ and $w^{ab}$, $\la_{ab}$ 
are the canonical ones. In other words, the second term in (\ref{wlOPE}) 
only contributes if ``$\beta = b$''.
Using the above formul\ae{} one can show that

\begin{eqnarray}
&& \!\!\!\!\!\!\!\!\!\!\!\!\!\! \!\!\!
w_\al \la^\al = w_+ \la^+ + w_a \la^a + \half w_{ab} \la^{ab} \,, 
\quad  w_\al (\ga^{ab})^\al{}_\bet \la^\bet = w^{ab} \la^+ 
+ \half \ep^{abcde} w_c \la_{de} \,, \label{wlid1}\\
&& \!\!\!\!\!\!\!\!\!\!\!\!\!\! \!\!\!
w_\al (\ga_{ab})^\al{}_\bet \la^\bet = -w_+ \la_{ab} -\half 
\ep_{abcde}w^{cd}\la^e  \label{wlid2} \,, \\ 
&& \!\!\!\!\!\!\!\!\!\!\!\!\!\! \!\!\!
w_\al (\ga^a{}_b)^\al{}_\bet \la^\bet = -w_b \la^a - w^{ac}\la_{cb} 
- \half\de^a_b (w_+ \la^+ - w_c\la^c +\half w^{cd}\la_{cd} ) \,. \label{wlid3}
\end{eqnarray}

\noindent
Using these results one can write down the expressions for the Lorentz scalar 
$\pa h = \frac{1}{2}w_\al \la^\al$ and the Lorentz currents 
$N_{mn} = \frac{1}{2}w_\al (\ga_{mn})^\al{}_\bet \la^\bet$ as

\begin{eqnarray}
&& \!\!\!\!\!\!\!\!\!\!\!\!\!\! \!\!\!
\partial h = -\half u_{ab} v^{ab} + \half \partial t +
{\ts \frac{3}{2} } \partial s \,, \qquad N^{ab} = -e^s v^{ab} \,, 
\label{hN1} \\
&& \!\!\!\!\!\!\!\!\!\!\!\!\!\! \!\!\! 
N_{ab} = e^{-s}(u_{ac}(v^{cd}u_{db})) + \half e^{-s}(u_{ab}(v^{cd}u_{cd})) 
-\half u_{ab}(e^{-s}(\pa t + \pa s)) +e^{-s}\pa u_{ab} \label{hN2} \,, \\
&& \!\!\!\!\!\!\!\!\!\!\!\!\!\! \!\!\!
\widetilde{N}^a{}_b = v^{ac}u_{cb} + {\ts \frac{1}{5}} \de^a_b v^{cd}u_{cd} \,, 
\qquad N = {\ts \frac{1}{\sqrt{5}} } (-{\ts \frac{5}{4} } (\pa t - \pa s) 
+ {\ts \frac{1}{4}} v^{cd} u_{cd}) \,. \label{hN3}
\end{eqnarray}

\noindent
Here $N^a{}_b$ has been decomposed into the traceless part, $\widetilde{N}^a{}_b$, 
and the trace part, $N$, according to $N^a{}_b = \widetilde{N}^a{}_b 
+ \frac{1}{\sqrt{5}} \de^a_b N$. 
There are normal ordering ambiguities in the above expressions.  
More precisely, the terms involving $\pa s$ and $\pa u$ are affected 
by normal ordering. The ambiguities are fixed by requiring that the OPE's 
have the right properties. In the above expressions we have indicated 
the normal ordering prescription by parentheses. 
The covariant OPE's obtained from the 
above expressions are given in section \ref{flat-back}.

Using the above formul\ae{} and $N_{mn} N^{mn} = 4[N_{ab}N^{ab} 
+ N^{ab} N_{ab} - 2 N^{a}{}_b N^b{}_a]$ one can show that the 
covariant expression for the stress tensor (\ref{covst}) 
reduces to the expression (\ref{ghosT})  
The derivation of this result makes extensive use of the following 
normal ordering rearrangement rules (here the parentheses indicate 
the normal ordering prescriptions)

\begin{eqnarray}
(A(B C)) - (B (A C)) &=& (([A,B]) C) \,, \\
((A B) C) - (B (A C)) &=& (A([C,B])) + (([C,A]) B) + ([(A B),C])\,.
\end{eqnarray}

\begingroup\raggedright\endgroup

\end{document}